\documentclass[%
 reprint,
 aps,
]{revtex4-1}

\usepackage{xcolor}
\usepackage{float}
\usepackage{graphicx}% Include figure files
\usepackage{dcolumn}% Align table columns on decimal point
\usepackage{bm}% bold math
\usepackage{braket}
\usepackage{amsmath}
\usepackage{float}
\usepackage{soul}
\usepackage{multirow}
\usepackage{xcolor}

\newcommand{\lp}{\left(}
\newcommand{\rp}{\right)}

\begin{document}

\title{Experimental quantum secret sharing with spin-orbit structured photons}

\author{Michael De Oliveira}
\author{Isaac Nape}%
\author{Jonathan Pinnell}%
\author{Najmeh TabeBordbar}%
\author{Andrew Forbes}%
\email{andrew.forbes@wits.ac.za}
\affiliation{%
School of Physics, University of the Witwatersrand, Johannesburg 2000, South Africa \\
}%


\begin{abstract}
\noindent Secret sharing allows three or more parties to share secret information which can only be decrypted through collaboration.  It complements quantum key distribution as a valuable resource for securely distributing information. Here we take advantage of hybrid spin and orbital angular momentum states to access a high dimensional encoding space, demonstrating a protocol that is easily scalable in both dimension and participants. To illustrate the versatility of our approach, we first demonstrate the protocol in two dimensions, extending the number of participants to ten, and then demonstrate the protocol in three dimensions with three participants, the highest realisation of participants and dimensions thus far.  We reconstruct secrets depicted as images with a fidelity of up to 0.979. Moreover, our scheme exploits the use of conventional linear optics to emulate the quantum gates needed for transitions between basis modes on a high dimensional Hilbert space with the potential of up to 1.225 bits of encoding capacity per transmitted photon. Our work offers a practical approach for sharing information across multiple parties, a crucial element of any quantum network. 
\end{abstract}

\maketitle
%%%%%%%%%%%%%%%%%%%%%%%%%%%%%%%%%%%%%%%%%%%%%%%%%%%%%%%%%%%%%%%%%%%%%%%%%%%%%%%%%%%%%%%%%%%%%%%%%%%%%%%%%%%%%%%%%%%%%%%%%%%%%%%%%%%%%%%%%%%%%%%%%%%%%%%%%%%%%%%%%%%%%%%%%%%%%%%%%%%%%%%%%%%%%%%%%%%%%%%%%%%%%%%%%%%%%%%%%%%%%%%%%%%%%%%%%%%%%%%%%%%%%%%%%%%%%%%%%%%%%%%%%%%%%%%%%%%%%%%%%%%%%%%%%%%%%%%%%%%%%%%%%%%%%%%%%%%%%%%%%%%%%%%%%%%%%%%%%%%%%%%%%%%%%%%%%%%%%%%%%%%%%%%%%%%%%%%%%%%%%%%%%%%%%%%%%%%%%%%%%%%%%%%%%%%%%%%%%%%%%%%%%%%%%%%%%%%%%%%%%%%%%%%%%%%%%%%%%%%%%%%%%%%%%%%%%%%%%
\section{\label{sec:introduction}Introduction}

\noindent In a world where cloud computing environments dominate our personal and corporate lives, secure communication and key distribution between multiple parties is a growing concern. This includes the secure sharing of encryption keys, missile launch codes, bank account information and social media profiles. In popular cryptography methods either a single copy of the encryption key is kept in one location for maximum secrecy or multiple copies of the same key are kept in different locations for greater reliability, but at an increased security risk. Secret sharing is a multiparty communication technique where a secret is divided and shared among $N$ parties and then securely reconstructed through collaboration, making it ideal for storing and sharing information that is highly sensitive, achieving both high levels of privacy and reliability \cite{ahlswede1993common,schneier1996applied}.

The first quantum secret sharing (QSS) scheme proposed the use of particle entangled states \cite{hillery1999quantum}.  In this protocol, three parties (Alice, Bob and Charlie) randomly choose between two measurement bases and independently measure their particle.  If their measurement results are correlated, Bob and Charlie can use their measurement bases and outcome information to determine the result of Alice's measurement, otherwise the round is discarded.  Since approximately half the instances will be discarded the intrinsic efficiency is about 50\%.   This protocol was improved to accommodate an arbitrary number of parties based on multi-particle qubit entanglement states \cite{sen2003unified}, and later to multi-particle $d$ dimensional entanglement states \cite{yu2008quantum}.  

Although much theoretical \cite{bandyopadhyay2000teleportation, nascimento2001improving, tyc2002share, karimipour2002entanglement, bagherinezhad2003quantum, xiao2004efficient, fu2004increasing, li2004multiparty,  han2008multiparty}, and (to a lesser extent) experimental \cite{tittel2001experimental, chen2005experimental, gaertner2007experimental} attention has focussed on QSS using multi-particle entangled states, progress has been limited by the intrinsic hurdle that the number of parties involved is bound by the number of entangled particles: this makes particle entanglement-based QSS inefficient and unscalable (multi-photon entanglement is notoriously inefficient). 

As a result of these limitations, two dimensional QSS schemes using single photon states, similar to those used in QKD, have been proposed \cite{guo2003quantum} and implemented \cite{schmid2005experimental}. Here, each party performs sequential unitary operations on the same particle instead of several entangled particles. The security was found to be less robust as compared to quantum key distribution (QKD) and susceptible to cheating strategies in that dishonest parties could infer some information about the choice of bases of another party \cite{he2007comment, qin2008special}.  To address this deficiency, multi-party high dimensional QSS protocols were theoretically proposed \cite{tavakoli2015secret,karimipour2015quantum,lin2016cryptanalysis,chen2018cryptanalysis} but with few suggestion as to how they might be (practically) implemented in the laboratory \cite{qin2019efficient, zhou2014implementation,smania2016experimental}. Challenges in high dimensional state preparation, transformation and detection, the key steps of any QSS protocol, have so far presented barriers to experimental realisation. 

Here we realise the first experimental high dimensional single photon QSS protocol using photons that are vectorially structured in their orbital angular momentum (OAM) and polarisation. Our approach requires only simple linear optical elements: spin-orbit coupling optics to prepare the initial state, half waveplates (HWPs) with dove prisms (DP) to encode the secret in the sequential phase transformation of each party, and a deterministic detector for all basis elements in the high dimensional vector space.  We successfully implement this protocol in two dimensions for ten parties, and three dimensions with three parties - the highest realisation of participants and dimensions thus far.  Our approach is scalable in the number of participants, highly efficient and provably secure.

%%%%%%%%%%%%%%%%%%%%%%%%%%%%%%%%%%%%%%%%%%%%%%%%%%%%%%%%%%%%%%%%%%%%%%%%%%%%%%%%%%%%%%%%%%%%%%%%%%%%%%%%%%%%%%%%%%%%%%%%%%%%%%%%%%%%%%%%%%%%%%%%%%%%%%%%%%%%%%%%%%%%%%%%%%%%%%%%%%%%%%%%%%%%%%%%%%%%%%%%%%%%%%%%%%%%%%%%%%%%%%%%%%%%%%%%%%%%%%%%%%%%%%%%%%%%%%%%%%%%%%%%%%%%%%%%%%%%%%%%%%%%%%%%%%%%%%%%%%%%%%%%%%%%%%%%%%%%%%%%%%%%%%%%%%%%%%%%%%%%%%%%%%%%%%%%%%%%%%%%%%%%%%%%%%%%%%%%%%%%%%%%%%%%%%%%%%%%%%%%%%%%%%%%%%%%%%%%%%%%%%%%%%%%%%%%%%%%%%%%%%%%%%%%%%%%%%%%%%%%%%%%%%%%%%%%%%%%%
\section{\label{sec:QSS}Single photon quantum secret sharing protocol}

\begin{figure*}[ht] 
	    \centering 
	    \includegraphics[width=\textwidth]{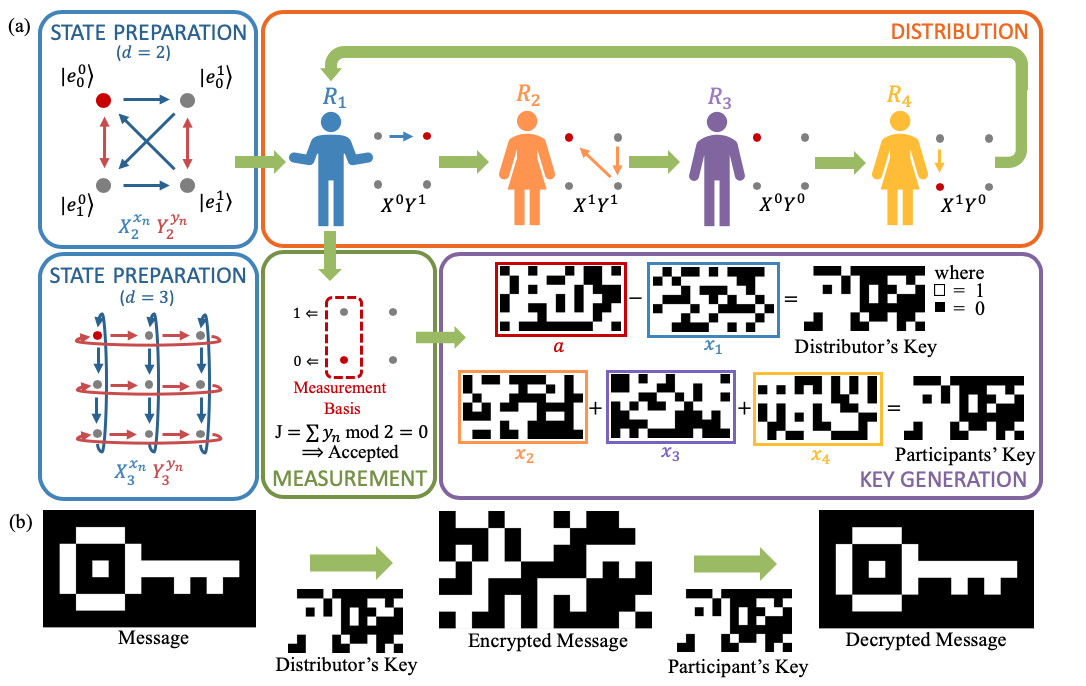}
	    \caption{\footnotesize (a) General scheme for a 4 party single qubit QSS scheme. The distributor, $R_{1}$, prepares an initial state from a set of $d=2$ MUBs. The qubit is then sequentially distributed to each party, who in turn performs a unitary phase operation given by $X_{d}^{x_{n}}Y_{d}^{y_{n}}$. The choice of $X^{x_{n}}$ is analogous to a change in local states within the basis, and a choice of $Y^{y_{n}}$ corresponds to a change of basis. The last participant sends the qubit back to the distributor. The distributor requests that parties $R_{2}, R_{3}, R_{4}$ broadcast their choice of $y_{n}$ and performs a measurement in a basis that leads to a deterministic result. The distributor can generate a secret key $x^{(scrt)}$ by using the measurement result to reset their choice of $x_{n}$. The other parties, upon collaborating and broadcasting their choice of $x_{n}$, can also generate the same secret key $x^{(scrt)}$. We also show the state preparation for $d=3$ dimensions. Note that the operators are cyclic in three dimensions because of the cyclic property of MUBs in odd prime dimensions. 
	    (b) The distributor can securely encrypt a message by applying a simple $XOR$ encryption operation using their generated secret key. The encrypted message, after being distributed, can be decrypted by each participant using their own secret key. At no point is the secret key shared among any participants.} 
	    \label{fig:ConceptFigure}
\end{figure*}

We begin by extending the single photon QSS protocol \cite{tavakoli2015secret} to prime dimensions and then we outline the general structure of a N-party QSS scheme using a single photon state. In this protocol multiple participants perform local operations on a single photon encoded in prime $d$ dimensions. Suppose a participant $R_{1}$, also known as the distributor, wants to share a secret key amongst  multiple parties, $R_{2}, \ldots, R_{N}$, then the QSS protocol can be summarised in four steps (see Fig. \ref{fig:ConceptFigure}):

\begin{enumerate}
	\item \textbf{State preparation:} The distributor, $R_{1}$, prepares an initial single photon state $\ket{e_{0}^{(0)}}$ from a set of mutual unbiased bases (MUB) in the desired prime dimension $d$. In our protocol, the MUBs are formulated from the logical basis, $\ket{\ell}$, as follows:
	\begin{align}
	    \ket{e_{k}^{(j)}}=\frac{1}{\sqrt{2}}\sum_{\ell=0}^{1}\omega^{\frac{1}{2}(j+2k)}\ket{\ell}
	\end{align}
	in two dimensions and in odd prime dimensions ($d'$) they are generalised as \cite{tavakoli2015secret},
	\begin{align}
	    \ket{e_{k}^{(j)}}=\frac{1}{\sqrt{d'}}\sum_{\ell=0}^{d'-1}\omega^{\ell(k +j \ell)}\ket{\ell} 
	\end{align} 
	where $k$ maps onto a mode from the $j^{th}$ MUB and  $\omega=\text{exp}( \frac{i2\pi}{d})$. Note that $\ell, j, k \in \{ 0, \ldots, d-1\}$.
	
	\item \textbf{Distribution:} The distributor modulates the photon initially in the state $\ket{e_{0}^{(0)}}$ with the operators $X_{d}^{x_{1}}Y_{d}^{y_{1}}$, where $x_{1}, y_{1}\in\{0,\ldots,d-1\}$ are chosen randomly and indicate how many times the operators should be applied. The photon is then sent sequentially to each participant $R_{2}, \ldots, R_{N}$, who upon receiving the single photon, randomly choose $x_{n}, y_{n}\in\{0,\ldots,d-1\}$, such that they apply the corresponding unitary operations $X_{d}^{x_{n}}Y_{d}^{y_{n}}$.
	
	To map between the MUB basis states, each party has access to two operators: $X_d$ and $Y_d$.  The operator $X_{d}$ is defined as,
	\begin{equation} 
	X_d=\sum^{d-1}_{\ell=0} \omega^\ell\ket{\ell}\bra{\ell}
	\label{eq:X_d}
	\end{equation}
	for prime dimensions. We adapted the protocol \cite{tavakoli2015secret} for two dimensions such that the operator $Y_d$ is defined as 
	\begin{equation} 
	Y_2 = \sum^{1}_{\ell=0} \omega^{\frac{1}{2}\ell}\ket{\ell}\bra{\ell}
	\label{eq:Y_2}
	\end{equation}
	in two dimensions and in odd prime dimensions ($d'$) as
	\begin{equation} 
	Y_{d'} = \sum^{d-1}_{\ell=0} \omega^{\ell^{2}}\ket{\ell}\bra{\ell}
	\label{eq:Y_d'}
	\end{equation} 
	The operator $X_{d}^{x_{n}}$ cycles through $x_{n}$ modes in the same basis, while the operator $Y_{d}^{y_{n}}$ cycles through $y_{n}$ MUBs, as shown in Fig. \ref{fig:ConceptFigure}(a). Using both operators in sequence results in the mapping between all MUB states which is crucial in the implementation of the single photon secret sharing protocol.   
	
	\item \textbf{Measurement:} After receiving the single photon from the last participant, the distributor request that parties $R_{2}, \ldots, R_{N}$ broadcast their choice of $y_{n}$ in a random order, keeping their value of $x_{n}$ a secret. By considering the sum of all $y_{n}$, the distributor chooses a measurement basis form the MUB set in such a way that the measurement leads to a deterministic result. In prime dimensions this is equivalent to applying the local unitary operator $Y_d^J$ and measuring the photon in the basis $\ket{e_{k}^{(J)}}$, where
    \begin{equation}
    J = \sum^{N}_{n=1}y_n \ \text{mod} \ d
    \label{eq:firstCheck1}
    \end{equation}
	The final measurement result obtained by the distributor is labelled $a\in\{0,\ldots,d-1\}$. Since the measurement is performed in a basis that yields a correlated result, the efficiency of the protocol is 100\% \cite{chen2018cryptanalysis}. 
	If Eq. \ref{eq:firstCheck1} holds, the participants have a strongly correlated selection of $x_n$,  satisfying
    \begin{equation}
        \sum^{N}_{n=1}x_n + C  =  a \ \text{mod} \  d
        \label{eq:SecCheck}.
    \end{equation}
    where we define $C = \lfloor \frac{1}{2}\sum^{N+1}_{n=1}y_n \rfloor$ for two dimensions, which accounts for the additional $X_2$ operator imparted by every odd number of $Y_2$ operators, and $C = 0$ in odd prime dimensions, due to the cyclic property of the operators in $d'$ dimensions. 
	
	\item \textbf{Key generation:} The distributor resets his value of $x_{1}^{(scrt)} = (a - x_1 + C) \ \text{mod} \ d$ according to the measurement result $a$. Consequently, if participants $R_{2}, \ldots, R_{N}$ collaborate and reveal among themselves their choice of $x_n$, they can reconstruct the distributors secret value $x_{1}^{(scrt)}=\sum_{n=2}^{N}x_{n} \ \text{mod} \ d$, which was previously only known to the distributor $R_{1}$. By repeating this procedure, the distributor can share a secret key among the rest $N-1$ participants. Using the secret key, the distributor can securely encrypt a message and distribute it to the participants, who in turn can use their own secret key to decrypt the message, as in Fig. \ref{fig:ConceptFigure}(b). 
\end{enumerate}

Participant $R_1$ checks the security, such that he randomly selects a subset of rounds. The degree of security specifications determines the size of the subset. In order to increase the security, as justified in \cite{chen2018cryptanalysis}, $R_1$ must make sure that the subset of valid rounds includes a round in which each participant broadcasts his choice of $y_{n}$ last. Each participant reveals their inferred value $x^{(scrt)}$ for the subset of rounds, which is compared to the value determined by the distributor. If there is a discrepancy any dishonest eavesdropping or cheating strategy is exposed.\\

In the next step, we investigate the necessary tools to implement a high-dimensional single photon QSS scheme. We explore vector modes and how we can implement unitary phase operators using simple linear optics.

%%%%%%%%%%%%%%%%%%%%%%%%%%%%%%%%%%%%%%%%%%%%%%%%%%%%%%%%%%%%%%%%%%%%%%%%%%%%%%%%%%%%%%%%%%%%%%%%%%%%%%%%%%%%%%%%%%%%%%%%%%%%%%%%%%%%%%%%%%%%%%%%%%%%%%%%%%%%%%%%%%%%%%%%%%%%%%%%%%%%%%%%%%%%%%%%%%%%%%%%%%%%%%%%%%%%%%%%%%%%%%%%%%%%%%%%%%%%%%%%%%%%%%%%%%%%%%%%%%%%%%%%%%%%%%%%%%%%%%%%%%%%%%%%%%%%%%%%%%%%%%%%%%%%%%%%%%%%%%%%%%%%%%%%%%%%%%%%%%%%%%%%%%%%%%%%%%%%%%%%%%%%%%%%%%%%%%%%%%%%%%%%%%%%%%%%%%%%%%%%%%%%%%%%%%%%%%%%%%%%%%%%
\section{Experimental realisation}

\begin{figure*}[ht]
		\centerline{\includegraphics[width=\linewidth]{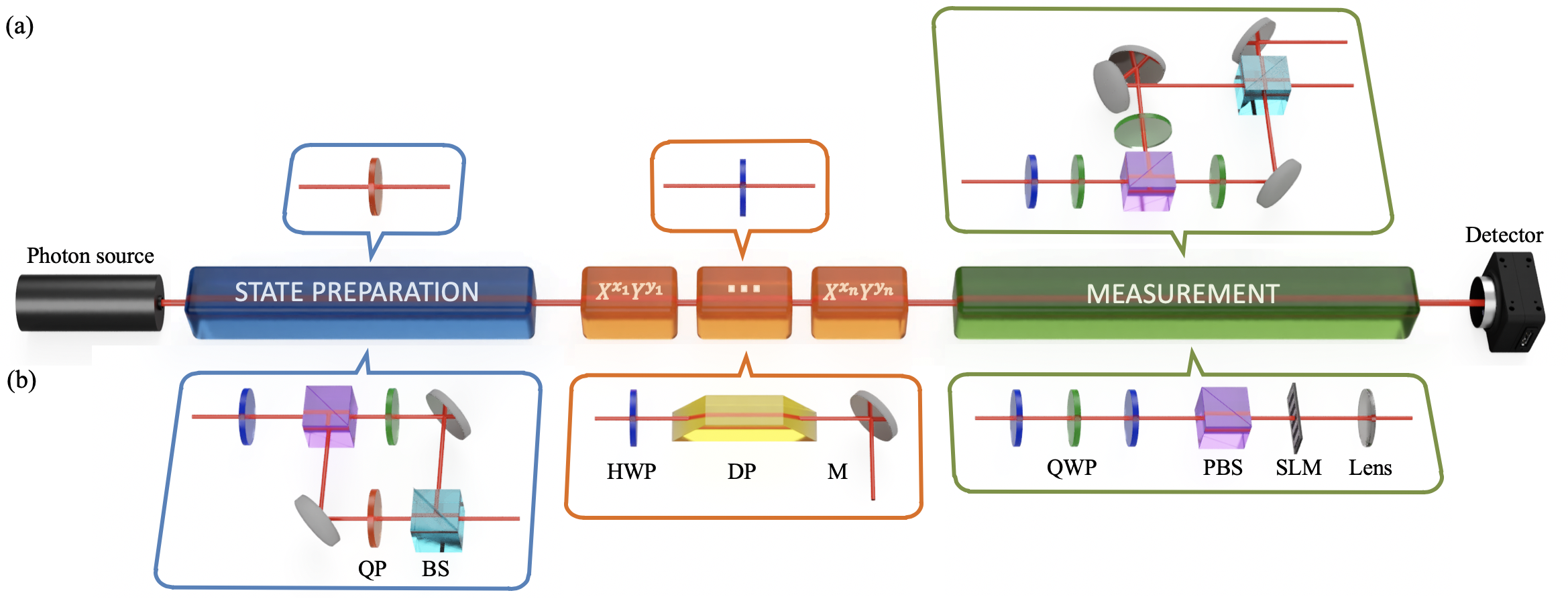}}
		\caption{\footnotesize Generalised experimental setup of a single photon quantum secret sharing scheme, showing the state preparation, distribution and measurement steps for (a) $d=2$ and (b) $d=3$ dimensions. The initial states are generated using a combination of geometric phase optics (i.e. a q-plate (QP)). The initial state is then sequentially communicated to each participant, who perform a unitary phase operator employed using simple linear optics such as a half waveplate (HWP) and a dove prism (DP). A HWP in the measurement step was used to perform the measurement in the same basis each time. The different states can be deterministically detected (a) using a combination of geometric phase control and multi-path interference using beam splitters (BM) and polarising beam splitter (PBS), or (b) via modal decomposition using a spatial light modulator (SLM). M are mirrors.}
		\label{fig:ComboSetup}
\end{figure*}

Here we introduce the tools (operations) needed for single photon secret sharing in  prime dimensions. Lastly, we show how the protocol can be implemented in both $d=2$ and $d=3$ dimensions using polarisation and OAM control (see Fig. \ref{fig:ComboSetup}).

%%%%%%%%%%%%%%%%%%%%%%%%%%%%%%%%%%%%%%%%%%%%%%%%%%%%%%%%%%%%%%%%%%%%%%%%%%%%%%%%%%%%%%%%%%%%%%%%%%%%%%%%%%%%%%%%%%%%%%%%%%%%%%%%%%%%%%%%%%%%%%%%%%%%%%%%%%%%%%%
\subsection{2-Dimensional realisation}

\begin{figure}[h]
	\centerline{\includegraphics[width=\linewidth]{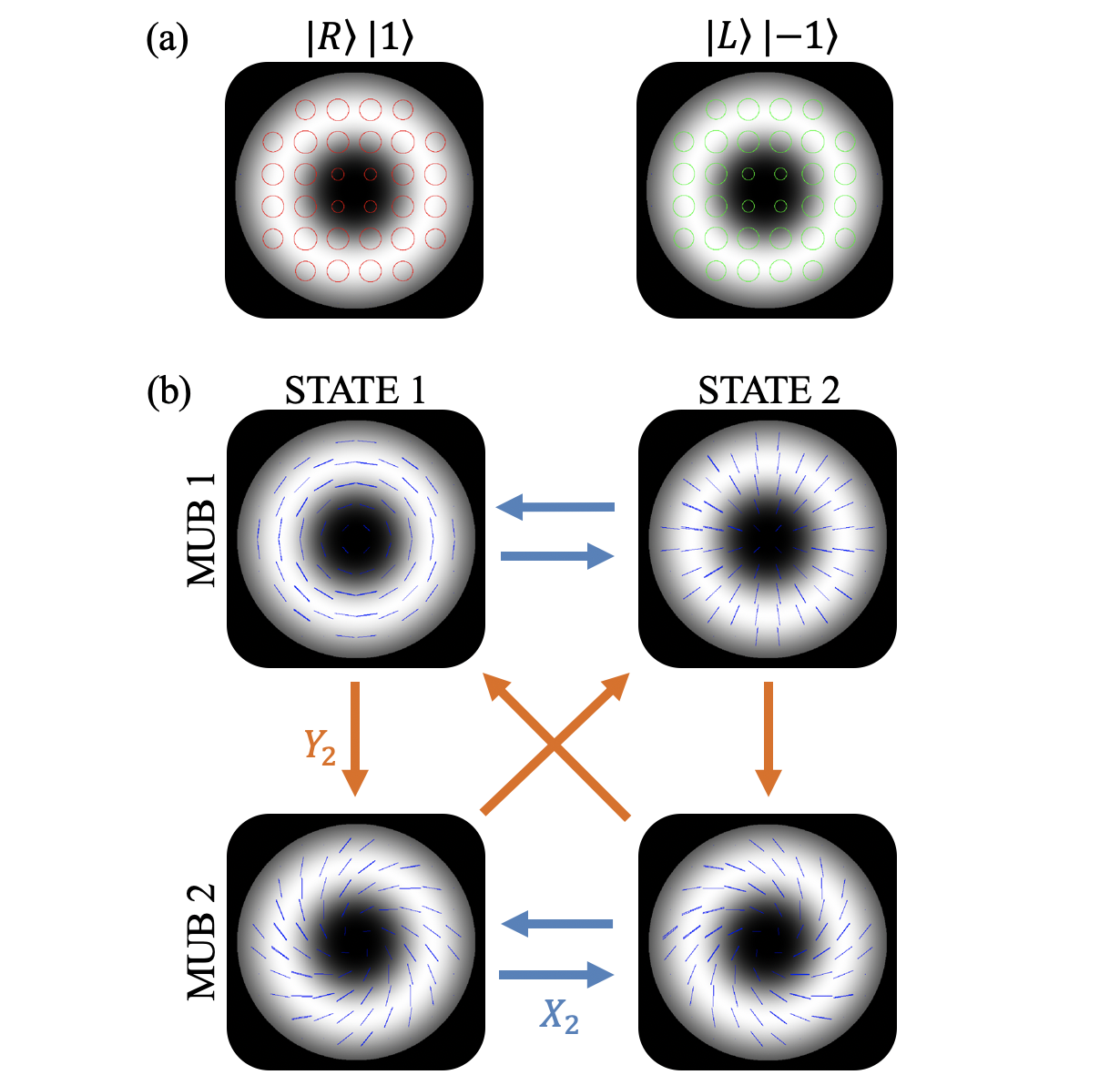}}
	\caption{\footnotesize (a) Illustration of the spin-orbit coupled modes that form our $d=2$ computational basis, from which we construct our MUBs. Right circularly polarised light is shown in red, left circularly polarised light is shown in green and linear polarisation in blue. (b) We realise the operators in $d=2$ using a half waveplate (HWP). A HWP at $\theta = \pi/4$ realises the $X_2$ operator, cycling between the states within the same basis; at $\theta = \pi/8$ we realise the $Y_2$ operator, moving between MUBs. Note that the $Y_2$ operator is not cyclic, due to the extra $X_2$ operator that is imparted by every odd number of $Y_2$.}
	\label{fig:states2D}
\end{figure}

If we consider the polarisation subspace coupled with the OAM subspace, spanned only by $|\ell|$, we can construct a two dimensional mode set, i.e $\mathcal{H}_2$=span(\{$\ket{R}\ket{\ell}, \ket{L}\ket{-\ell}\}$) as illustrated in Fig. \ref{fig:states2D}(a). The basis states can be mapped as orthogonal column vectors,

\begin{align}
\ket{R}\ket{\ell}=\begin{pmatrix}
      1\\0
     \end{pmatrix}, \ \ket{L}\ket{-\ell}=\begin{pmatrix}
      0\\1
     \end{pmatrix} \label{eq:2Dstates}
\end{align}
This allows us to map the MUBs (see Fig. \ref{fig:states2D}(b)) as row vectors in matrix form as follows:
\begin{equation}
\text{M}_1=\frac{1}{\sqrt{2}}\begin{pmatrix}
      1&1\\
      1&-1\\
     \end{pmatrix}, \  \text{M}_2=\frac{1}{\sqrt{2}}\begin{pmatrix}
      1&i\\
      1&-i\\
     \end{pmatrix} \label{eq:2Dbasis}
\end{equation}

The first step in implementing the protocol is preparing the photon in the initial state within our MUB set. We generated the initial state, $\ket{e_{0}^{(0)}}=\frac{1}{\sqrt{2}}\lp\ket{R}\ket{\ell}+\ket{L}\ket{-\ell}\rp$ denoted by $\ket{\Psi_{0}}$, from a horizontally polarised Gaussian beam incident on a spin-orbit coupling $q$-plate \cite{marrucci2006optical, marrucci2011spin}.

The next step is to find a way to independently move between each MUB state, by applying the required operators. This is easily implemented by a half waveplate (HWP). It is straight forward to see that a HWP acting on the initial state $\ket{e_{0}^{(0)}}$, induces a relative phase difference, $e^{i4\theta}$, between the circular polarisation states. This can be summarized as
    \begin{equation}
        \hat{U}(\theta) \propto \begin{pmatrix}
         1  &   0   \\
         0   &   e^{i4\theta}\\
         \end{pmatrix}
    \end{equation}
where $\theta\in\{0, \pi/8, \pi/4, 3\pi/8\}$ is the rotation angle of the HWP, corresponding to the transformations $\hat{U}(\theta) = \{X_{2}^{0}Y_{2}^{0}, X_{2}^{0}Y_{2}^{1}, X_{2}^{1}Y_{2}^{0}, X_{2}^{1}Y_{2}^{1}\}$. In this way, Fig. \ref{fig:states2D}(b) shows that we can move independently between all MUBs.

Once the initial state is sent through a set of even N consecutive HWPs, allowing each party to apply their unitary operator, the final state of the photon will be:
    \begin{equation}
        \ket{\Psi_{N}}=\frac{e^{i\Omega}}{\sqrt{2}}\left[ \ket{R}\ket{\ell}+e^{i\Phi}\ket{L}\ket{-\ell}\right]
    \end{equation}
where $\Omega =\lp-i\rp^{N}e^{-2i\lp\sum_{n=1}^{N}\lp-1\rp^{n+1}\theta_{n}\rp}$ and $\Phi=4\sum_{n=1}^{N}\lp -1\rp^{n+1}\theta_{n}$. The distributor then applies the corresponding operator for $\phi_{J}\in\{0, \pi/2\}$ using a HWP, such that performing the measurement in the basis $\frac{1}{\sqrt{2}}\lp\ket{R}\ket{\ell}+e^{i\phi_{J}}\ket{L}\ket{-\ell}\rp$ leads to deterministic result.
    
Next, we discuss the detection system used to distinguish between all MUB states. The different states can be deterministically detected using a combination of geometric phase control and multi-path interference as seen in Fig. \ref{fig:ComboSetup}(a). The beam was split into two polarisation dependent paths using a combination of quarter waveplates (QWP) and a polarizing beam splitter (PBS), such that the state of the qubit becomes,
	\begin{equation}
		\ket{\Psi_{N}}=\frac{e^{i\Omega}}{\sqrt{2}}[\ket{R}_{a}\ket{1} _{a} + e^{i\Phi}\ket{L}_{b}\ket{-1}_{b}]
	\end{equation}
where the subscripts $a$ and $b$ refer to the polarisation dependent paths. The photon paths were interfered at a 50:50 beam splitter (BS), setting the dynamic phase difference between the two paths to $\pi/2$. An extra reflection was added to one path so that the number of reflections, and thus the polarisation of the two output paths, was automatically reconciled. Henceforth, we will drop the polarisation kets in the expression as the polarisation information is path dependent. The resulting state after the BS is
	\begin{equation}
		\ket{\Psi_{N}'}=\frac{e^{i\Omega}}{2}[(1-e^{i\Phi})\ket{1} _{c} + i (1+e^{i\Phi})\ket{-1}_{d}]
	\end{equation}
where the subscript $c$ and $d$ refer to the output paths of the beam splitter. From this equation we see that the detection scheme is in fact deterministic for given  values of $\Phi$, such that all the light will be in either path $c$ or $d$.

Next, we extend the two dimensional implementation to three dimensions, using a similar linear optics setup.

%%%%%%%%%%%%%%%%%%%%%%%%%%%%%%%%%%%%%%%%%%%%%%%%%%%%%%%%%%%%%%%%%%%%%%%%%%%%%%%%%%%%%%%%%%%%%%%%%%%%%%%%%%%%%%%%%%%%%%%%%%%%%%%%%%%%%%%%%%%%%%%%%%%%%%%%%%%%%%%
\subsection{3-Dimensional realisation}

\begin{figure}[h]
	\centerline{\includegraphics[width=\linewidth]{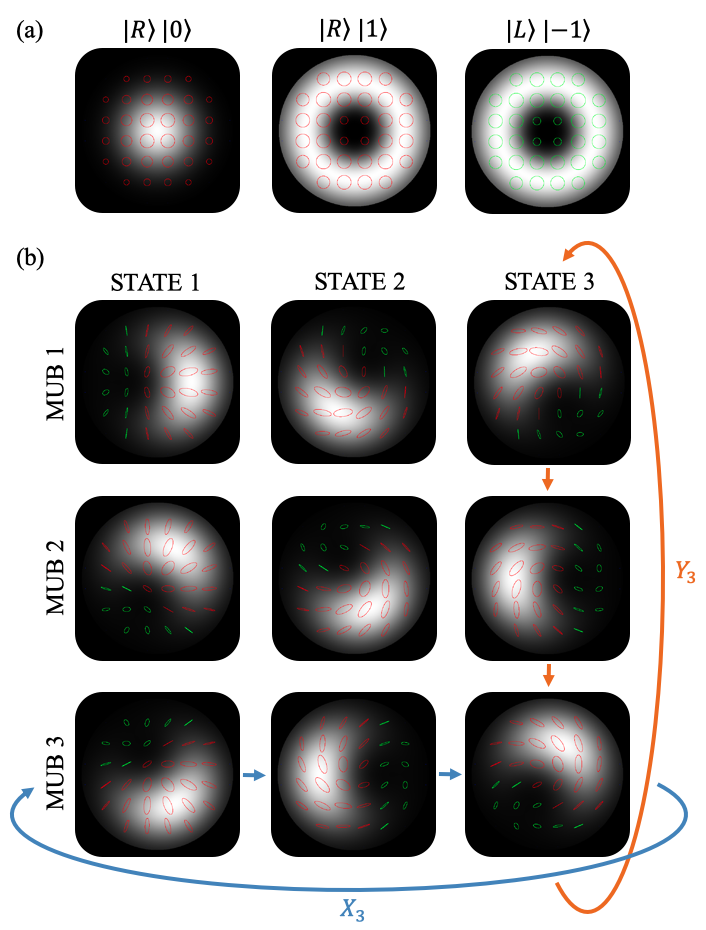}}
	\caption{\footnotesize (a) Illustration of the spin-orbit coupled modes that form our $d=3$ computational basis, from which we construct our MUBs. Right circularly polarised light is shown in red and left circularly polarised light is shown in green.  (b) Here we show the cyclic nature of the operators in $d=3$. A dove prism (DP) allows us to realise the $X_3$ gate, cycling between the states within the same basis, and a half waveplate (HWP) allows us to realise the $Y_3$ gate, cycling between the MUBs.}
	\label{fig:states3D}
\end{figure}

We now consider a mode set that spans a three dimensional (qutrit) space of spin-orbit coupled modes, i.e $\mathcal{H}_3$=span(\{$\ \ket{R}\ket{0}), \ket{R}\ket{\ell}, \ket{L}\ket{-\ell}\}$) as depicted in Fig. \ref{fig:states3D}(a).  If we map the basis states as orthogonal column vectors, i.e, 
\begin{align}
\ket{R}\ket{0}=\begin{pmatrix}
      1\\0\\0
     \end{pmatrix}, \ket{R}\ket{\ell}=\begin{pmatrix}
      0\\1\\0
     \end{pmatrix},\ket{L}\ket{-\ell}=\begin{pmatrix}
      0\\0\\1
     \end{pmatrix} \label{eq:3Dbasis}
\end{align}
the the MUBs can be mapped as row vectors in matrix form, where $\omega=\text{exp}( \frac{i2\pi}{3})$, as follows
\begin{equation}
    \text{M}_1=\frac{1}{\sqrt{3}}\begin{pmatrix}
      1&1&1\\
      1&\omega&\omega^2\\
      1&\omega^2&\omega\\
     \end{pmatrix}, \ \text{M}_2=\frac{1}{\sqrt{3}}\begin{pmatrix}
      1&1&\omega\\
      1&\omega&1\\
      \omega&1&1\\
     \end{pmatrix}, \nonumber
\end{equation}
\begin{equation}
    \text{M}_3=\frac{1}{\sqrt{3}}\begin{pmatrix}
      1&1&\omega^2\\
      1&\omega^2&1\\
      \omega^2&1&1\\
     \end{pmatrix}
\end{equation}

The initial state was prepared using a interferometric combination of a q-plate, HWP and beam splitter, as in Fig. \ref{fig:ComboSetup}(b). We further engineer the required operators by using a half waveplate in combination with a dove-prism as illustrated. As before, the HWP induces a relative phase difference, $e^{4i\theta}$, between the circular polarisation DoF and the DP imparts a phase which proportional to the OAM state. A mirror after the dove prism is needed to invert the final OAM state. The unitary transformation, in the basis from Eq. \ref{eq:3Dbasis} can be summarised as
\begin{equation}
\hat{U}(\theta, \gamma) \propto \begin{pmatrix}
     1  &   0 & 0   \\
     0  &   e^{-i\gamma\ell_2}     &   0\\
     0  &   0   &   e^{-i(\gamma\ell_3-4\theta)}\\
     \end{pmatrix},
\end{equation}
where $\theta\in\{0, \pi/6, 2\pi/6\}$ is the rotation angle of the HWP and $\gamma\in\{0, \pi/3, 2\pi/3\}$ is the rotation angle of the DP. The DP allows us to realise the $X_3$ gate, cycling between the states within the same basis, and the HWP allows us to realise the $Y_3$ gate, cycling between the MUBs (see Fig. \ref{fig:states3D}(b)).\\

The detection system included mapping our vector basis to a scalar basis using a set of half waveplate and quarter waveplates. We measured the detection probabilities of each MUB state using match filters encoded on the SLM via modal decomposition (see supplementary material). The detection modes where encoded as phase and amplitude holograms \cite{arrizon2007pixelated} on a Holoeye Pluto spatial light modulator (SLM) - a well established technique for spatial mode detection \cite{forbes2016creation}. 

However, using a quantum Fourier transform (QFT) to map between the MUB superpositions of OAM modes to the OAM standard basis, one can  deterministicaly sort the MUBs and thereafter sort the OAM modes. In three dimensions, a QFT for OAM has been proposed \cite{jo2019efficient}. The technique exploits the tritter \cite{zukowski1997realizable}, by using path and phase control. Once the mapping between the MUB and OAM basis is achieved, mode sorters can be used deterministicly to measure the OAM modes \cite{ berkhout2010efficient}. Mode sorting has been extensively used for both scalar \cite{mirhosseini2013efficient} and vector modes \cite{ndagano2017deterministic}.	
%%%%%%%%%%%%%%%%%%%%%%%%%%%%%%%%%%%%%%%%%%%%%%%%%%%%%%%%%%%%%%%%%%%%%%%%%%%%%%%%%%%%%%%%%%%%%%%%%%%%%%%%%%%%%%%%%%%%%%%%%%%%%%%%%%%%%%%%%%%%%%%%%%%%%%%%%%%%%%%%%%%%%%%%%%%%%%%%%%%%%%%%%%%%%%%%%%%%%%%%%%%%%%%%%%%%%%%%%%%%%%%%%%%%%%%%%%%%%%%%%%%%%%%%%%%%%%%%%%%%%%%%%%%%%%%%%%%%%%%%%%%%%%%%%%%%%%%%%%%%%%%%%%%%%%%%%%%%%%%%%%%%%%%%%%%%%%%%%%%%%%%%%%%%%%%%%%%%%%%%%%%%%%%%%%%%%%%%%%%%%%%%%%%%%%%%%%%%%%%%%%%%%%%%%%%%%%%%%%%%%%%%
\section{\label{sec:Results}Results}

Here we present the results for our implementation of the quantum secret sharing protocol with single photon states in $d=2$ and $d=3$ dimensions. For practical purposes, the experiment was first performed with a classical light source and a ccd camera. Later, the light source was attenuated to an average photon number of $\mu= 0.02$ per pulse. Although weak coherent states cannot be used without photon splitting strategies this could, in principle, be overcome by preparing and testing the transmission properties of some decoy states. In the single photon regime, the measurement system includes coupling the photons through fibres to avalanche photon detectors (APD). 
 
%%%%%%%%%%%%%%%%%%%%%%%%%%%%%%%%%%%%%%%%%%%%%%%%%%%%%%%%%%%%%%%%%%%%%%%%%%%%%%%%%%%%%%%%%%%%%%%%%%%%%%%%%%%%%%%%%%%%%%%%%%%%%%%%%%%%%%%%%%%%%%%%%%%%%%%%%%%%%%%
\subsection{2-Dimensional results}

\begin{figure}[h]
	\centerline{\includegraphics[width=\linewidth]{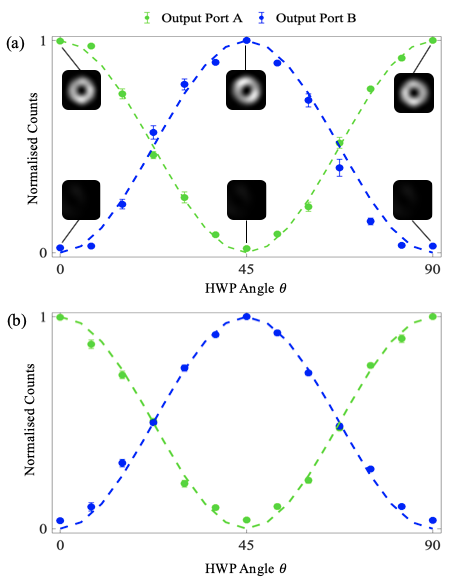}}
	\caption{\footnotesize Detection of superposition of vector states. Each graph shows the detection (normalized intensity) of the photons in a superposition of the vector states $\ket{\Psi'_{N}}$, generated by rotating the HWP angle $\theta$, using (a) ccd camera and (b) photodiodes in the single photon regime.  Each data point was generated by averaging over 35 measurements. The dashed lines show the theoretical curve.}
	\label{fig:interferomenterchracterisation}
\end{figure}

The two dimensional detection results of our vector basis ares shown in Fig. \ref{fig:interferomenterchracterisation}. This was performed by rotating the angle $\theta$ of the HWP and measuring the intensity of each output port using a ccd camera at each port (see Fig. \ref{fig:interferomenterchracterisation}(a)) and in the single photon regime, using single photon detectors (see Fig. \ref{fig:interferomenterchracterisation}(b)). 

\begin{figure*}[ht]
	\centerline{\includegraphics[width=\textwidth]{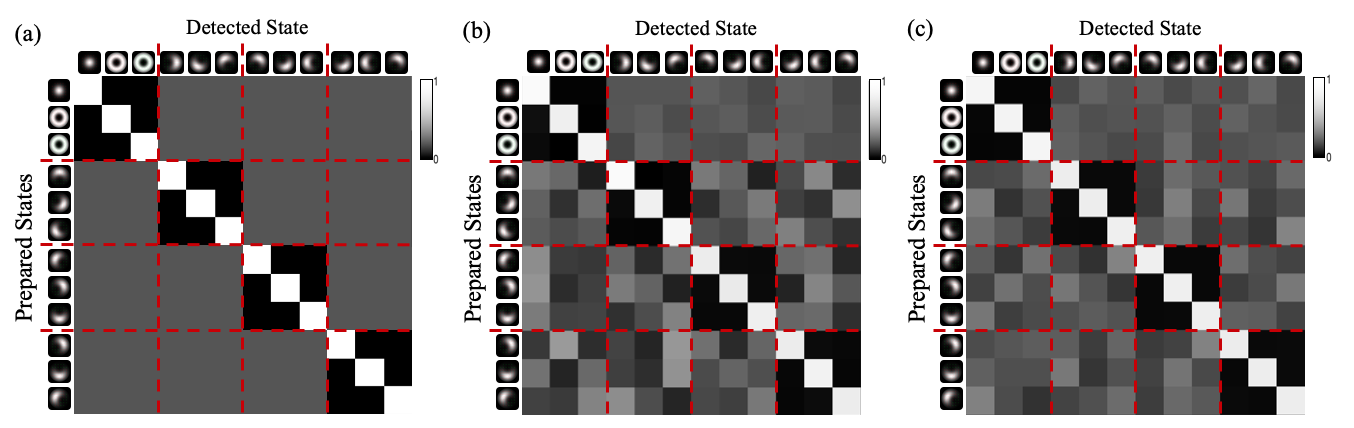}}
	\caption{\footnotesize Crosstalk matrices shown theoretically in (a) and experimentally in (b)
and (c), for classical light and the single photon regime respectively. This shows the scattering probabilities for modes prepared and detected in identical bases (diagonal) and the overlap between modes from mutually unbiased bases (off diagonal).}
	\label{fig:ProbMatrix}
\end{figure*}

There is an excellent agreement between the experimental results (data points) and the theory (dashed curves). The Visibility, $V$, of the detection scheme in each output port was calculated using the equation:
	\begin{equation}
		V = \frac{|\mathcal{I}_{max}-\mathcal{I}_{min}|}{\mathcal{I}_{max}+\mathcal{I}_{min}}
	\end{equation}
where $\mathcal{I}$ is the intensity in each arm. Spatial filtering was applied to the data obtained using the ccd camera to remove unwanted noise, resulting in $V = 0.958 \pm 0.005$. In our system, the errors are introduced by the additive imperfections in the half waveplates causing slight misalignment in the setup. The visibility for the single photon regime was measured to be $V = 0.924 \pm 0.003$, which can be accounted for by the photon loss in fibre coupling and detector dark counts. Nonetheless, such values imply the use of a well-aligned and stable interferometer. 

For phase-coding setups, the fidelity of the detection system is related to the interference visibility by \cite{gisin2002quantum},
\begin{equation}
    F = \frac{1+V}{2}
    \label{eq:fidelity}
\end{equation}

 Hence, the fidelity of the system was calculated to be $F = 0.979 \pm 0.005$ for the classical implementation and $F = 0.962 \pm 0.003$ for the single photon regime. Using this deterministic detector, we can detect any arbitrary superposition of our vector basis with high fidelity.

%%%%%%%%%%%%%%%%%%%%%%%%%%%%%%%%%%%%%%%%%%%%%%%%%%%%%%%%%%%%%%%%%%%%%%%%%%%%%%%%%%%%%%%%%%%%%%%%%%%%%%%%%%%%%%%%%%%%%%%%%%%%%%%%%%%%%%%%%%%%%%%%%%%%%%%%%%%%%%%
\subsection{3-Dimensional results}

To demonstrate the feasibility of our secret sharing scheme in three dimensions, we verify that the d+1 MUBs are each orthogonal with respect to each other by measuring the scattering probabilities. The crosstalk matrix is shown theoretically in Fig.~\ref{fig:ProbMatrix} (a) and experimentally in Fig.~\ref{fig:ProbMatrix} (b) and Fig.~\ref{fig:ProbMatrix} (c), for the classical and single photon regime respectively. To obtain the results we first prepared the initial superposition state $\ket{e_{0}^{(0)}}$ and applied the $X_3$ and $Y_3$ gates to iterate through the various basis modes and MUB mode sets. Using a set of waveplates, we mapped the circular polarisation photon states to the horizontal polarisation state and performed projective measurements via modal decomposition.

From the crosstalk matrices, we measured an average fidelity of $F = 0.946 \pm 0.003$ when using classical light and similarly we measured $F = 0.938 \pm 0.001$ in the single photon regime, which is $F = 1$ for a perfect system. In our system, the errors are introduced by imperfections, including the rotation of the dove prism and half waveplates causing slight misalignment in the setup.

%%%%%%%%%%%%%%%%%%%%%%%%%%%%%%%%%%%%%%%%%%%%%%%%%%%%%%%%%%%%%%%%%%%%%%%%%%%%%%%%%%%%%%%%%%%%%%%%%%%%%%%%%%%%%%%%%%%%%%%%%%%%%%%%%%%%%%%%%%%%%%%%%%%%%%%%%%%%%%%
\subsection{Security analysis}

From the measured detection fidelities, we performed a security analysis on our QSS scheme for $d=2$ and $d=3$ dimensions. The results of the analysis are summarised in Table \ref{table:results}.

The quantum bit error rate (QBER), reflecting the probability of making detection errors, is related to the fidelity by,
\begin{equation}
    \mbox{QBER} = 1 - F
    \label{eq:QBER}
\end{equation}
which is 0 for a perfect system. The detection fidelities translated into an optical QBER between $0.021$ and $0.062$, well below the $0.110$ and $0.156$ bounds for unconditional security against coherent attacks in two and three dimensions respectively \cite{cerf2002security}. \\
 
\begin{table}[h]
	\centering
	\begin{tabular}{c c c c c}
	\hline
	\multirow{2}{*}{\textbf{Measures }} & \multicolumn{2}{c}{\textbf{d=2}} & \multicolumn{2}{c}{\textbf{d=3}} \\ %\cline{2-5}
    & \textbf{ Classical} & \textbf{Quantum } & \textbf{ Classical} & \textbf{Quantum } \\ \hline \hline
	$F$ & 0.979	&0.958 & 0.946	& 0.938 \\
    $QBER$ & 0.021 & 0.038 & 0.054	& 0.062\\
    $I$ & 0.853	& 0.767 & 1.225	& 1.187 \\
    \hline
	\end{tabular}
	\caption{Summary of the $d=2$ and $d=3$ experimental results for our secret sharing protocol, for both the classical regime using the CCD camera as a detector and for the single photon regime using APDs. We show the experimental values of the detection fidelity (F), the quantum bit error rate (QBER) in bits per photon and mutual information (I) between distributor and participants.}
   	\label{table:results}
	\end{table}
	
\begin{figure*}[ht]
	\centerline{\includegraphics[width=\textwidth]{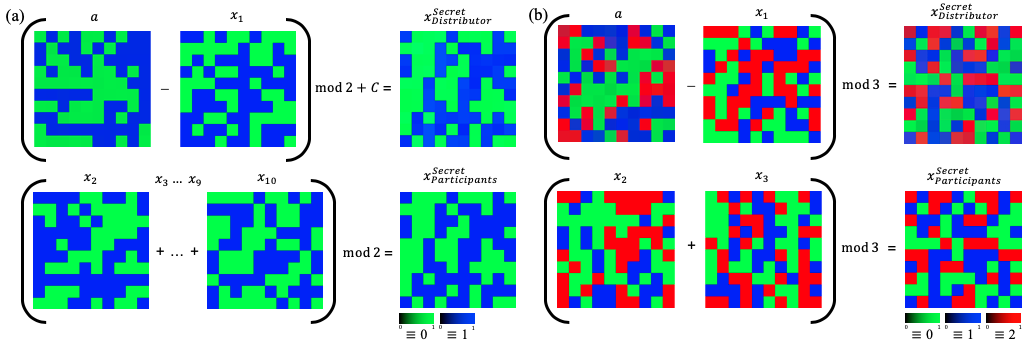}}
	\caption{\footnotesize Experimentally generated distributor's and participants' secret keys, in (a) $d=2$  and (b) $d=3$ dimensions, by implementing the protocol for 100 valid runs. The colour bars indicate the measured probability of generating a 0, 1 or 2.}
	\label{fig:GeneratedKeys}
\end{figure*}

From the fidelity we can calculate the mutual information, $I$x. This places a bound on the amount of information that can be shared between the distributor and participants. This bound is only due to the generation and detection fidelities, and not intrinsic to the protocol itself. This is given by,
\begin{equation}
    I = \log_{2}(d) + F\log_{2}(F) + (1-F)\log_{2}(\frac{1-F}{d-1}).
    \label{eq:mutualinfo}
\end{equation}

For a perfect system we would expect a value of 1 bit per photon in a $d=2$ qubit system and 1.58 bits per photon in in a $d=3$ qutrit system. For $d=3$ this was measured to be nearly $1.5\times$ the maximum achievable in $d=2$ dimensions. We note that increasing the dimension of the quantum secret sharing protocol, did result in higher mutual information capacity.

%%%%%%%%%%%%%%%%%%%%%%%%%%%%%%%%%%%%%%%%%%%%%%%%%%%%%%%%%%%%%%%%%%%%%%%%%%%%%%%%%%%%%%%%%%%%%%%%%%%%%%%%%%%%%%%%%%%%%%%%%%%%%%%%%%%%%%%%%%%%%%%%%%%%%%%%%%%%%%%
\subsection{Secret key generation:}

To corroborate the advantage of our protocol utilising a higher dimensional encoding space, we experimentally shared a secret in both $d=2$ and $d=3$ dimensions using the experimental setups described. 

In two dimensions, the protocol was performed by $N = 10$ participants - the highest number of participants realised thus far - each equipped with a $X_2$ and $Y_2$ gate (half waveplate). We ran the protocol for 100 valid runs, resulting in a generated secret key of 100 bits. The  results  are  shown  in  Fig. \ref{fig:GeneratedKeys}(a), for the identical secret key retrieved by the distributor and shared between the participants. The distributor's secret key was determined by resetting his choice of $x_{1}$ using the measurement results and the participants choice of $y_n$. The participants shared secret key was calculated by summing the keys of the participants $R_2, \cdots, R_{10}$, modulus 2. By performing the measurement in a basis that would yield correlated results (see Ref. \cite{chen2018cryptanalysis}), we successfully implemented the two dimensional protocol with an efficiency of 100\%.\\

Next, exploiting the higher dimensional ($d=3$) encoding space, we shared a secret key between $N = 3$ participants, each equipped with the $X_3$ gate (dove prism) and $Y_3$ gate (half waveplate). The results are shown in Fig. \ref{fig:GeneratedKeys}(b), for the secret code retrieved by the distributor and shared between the participants. The keys are identical as desired. Using the high dimensional protocol for 100 valid runs we generated a secure key that was 158 bits.

%%%%%%%%%%%%%%%%%%%%%%%%%%%%%%%%%%%%%%%%%%%%%%%%%%%%%%%%%%%%%%%%%%%%%%%%%%%%%%%%%%%%%%%%%%%%%%%%%%%%%%%%%%%%%%%%%%%%%%%%%%%%%%%%%%%%%%%%%%%%%%%%%%%%%%%%%%%%%%%%%%%%%%%%%%%%%%%%%%%%%%%%%%%%%%%%%%%%%%%%%%%%%%%%%%%%%%%%%%%%%%%%%%%%%%%%%%%%%%%%%%%%%%%%%%%%%%%%%%%%%%%%%%%%%%%%%%%%%%%%%%%%%%%%%%%%%%%%%%%%%%%%%%%%%%%%%%%%%%%%%%%%%%%%%%%%%%%%%%%%%%%%%%%%%%%%%%%%%%%%%%%%%%%%%%%%%%%%%%%%%%%%%%%%%%%%%%%%%%%%%%%%%%%%%%%%%%%%%%%%%%%%%%%%%%%%%%%%%%%%%%%%%%%%%%%%%%%%%%%%%%%%%%%%%%%%%%%%%
\section{\label{sec:discussion}Discussion}
Transverse spatial modes of light carrying orbital angular momentum have become ubiquitous for encoding quantum information with promising applications in quantum communication. Spanning the $d\geq2$ dimensional Hilbert space, OAM modes have proven invaluable for secure and robust communication, and thus have the potential to increase the mutual information and security of quantum channels in QSS. However, despite its many potential advantages, the complete realization of a high dimensional quantum cryptography with OAM, so far, has been limited by technical difficulties arising in the full manipulation and transmission of this degree of freedom (DoF). 

To overcome these restraints, photon states encoded in different DoFs, called hybrid entangled states, have attracted a lot of attention \cite{nagali2010experimental}. Spin-orbit coupled states, e.g. vector modes, have been used to complete the entanglement purification in photon pairs for polarization Bell states \cite{sheng2010deterministic, li2010deterministic}. Similar they have been used to overcome the limiting channel capacity of superdense coding \cite{barreiro2008beating} and to realise a high capacity QKD protocol \cite{wang2009doubling}. 

We have reported a novel scheme for sharing secure keys between multiple parties by interfacing different DoF, namely spin and orbital angular momentum of single photons in high-dimensions ($d=3$). Our scheme can be extended to multiple participants and requires conventional linear optical elements making it easily scalable. For a practical implementation waveplates and dove prisms can be rotated using electronically driven rotation mounts \cite{toninelli2019concepts}, whose rotation rate would be the only limiting factor with regards to the generation rates. The spatial modes used are OAM modes of light, which can be represented by LG modes and thus are the natural modes of quadratic media. Moreover, the scheme can be exploited over long distances (up to 1 km kilometer \cite{cui2017distribution}) using few mode fibers, since our basis modes lie in the first two mode groups which may have low group delays and minimal crosstalk, if chosen carefully. Applications can also be extended to underwater channels, although the main challenge would be overcoming deleterious effects, like turbulence which could reduce the QBER as previously shown for QKD. 

%%%%%%%%%%%%%%%%%%%%%%%%%%%%%%%%%%%%%%%%%%%%%%%%%%%%%%%%%%%%%%%%%%%%%%%%%%%%%%%%%%%%%%%%%%%%%%%%%%%%%%%%%%%%%%%%%%%%%%%%%%%%%%%%%%%%%%%%%%%%%%%%%%%%%%%%%%%%%%%%%%%%%%%%%%%%%%%%%%%%%%%%%%%%%%%%%%%%%%%%%%%%%%%%%%%%%%%%%%%%%%%%%%%%%%%%%%%%%%%%%%%%%%%%%%%%%%%%%%%%%%%%%%%%%%%%%%%%%%%%%%%%%%%%%%%%%%%%%%%%%%%%%%%%%%%%%%%%%%%%%%%%%%%%%%%%%%%%%%%%%%%%%%%%%%%%%%%%%%%%%%%%%%%%%%%%%%%%%%%%%%%%%%%%%%%%%%%%%%%%%%%%%%%%%%%%%%%%%%%%%%%%%%%%%%%%%%%%%%%%%%%%%%%%%%%%%%%%%%%%%%%%%%%%%%%%%%%%%
\section{\label{sec:conclusion}Conclusion}
In conclusion, we successfully implemented two-dimensional single photon QSS for 10 parties, the highest realisation so far. We further extended our scheme to higher dimensions by interfacing independent degrees of freedom, providing a natural extension to high-dimensional QSS. Our approach shows that by using hybrid polarisation and OAM encoding, it is possible to realise a $d=2$ and $d=3$ dimensional single photon QSS using conventional linear optical elements. Further, by exploiting the non-separability of polarization and OAM in our choice of spatial modes, we were able to realise transitions on a high dimensional  Hilbert space, mapping between different MUB states, demonstrating the advantage of interfacing independent DoF. Our practical scheme is scalable to an unlimited number of participants and can be realised using current technologies.

%%%%%%%%%%%%%%%%%%%%%%%%%%%%%%%%%%%%%%%%%%%%%%%%%%%%%%%%%%%%%%%%%%%%%%%%%%%%%%%%%%%%%%%%%%%%%%%%%%%%%%%%%%%%%%%%%%%%%%%%%%%%%%%%%%%%%%%%%%%%%%%%%%%%%%%%%%%%%%%
\bibliography{SSSINotes.bib}

\begin{thebibliography}{48}%
\makeatletter
\providecommand \@ifxundefined [1]{%
 \@ifx{#1\undefined}
}%
\providecommand \@ifnum [1]{%
 \ifnum #1\expandafter \@firstoftwo
 \else \expandafter \@secondoftwo
 \fi
}%
\providecommand \@ifx [1]{%
 \ifx #1\expandafter \@firstoftwo
 \else \expandafter \@secondoftwo
 \fi
}%
\providecommand \natexlab [1]{#1}%
\providecommand \enquote  [1]{``#1''}%
\providecommand \bibnamefont  [1]{#1}%
\providecommand \bibfnamefont [1]{#1}%
\providecommand \citenamefont [1]{#1}%
\providecommand \href@noop [0]{\@secondoftwo}%
\providecommand \href [0]{\begingroup \@sanitize@url \@href}%
\providecommand \@href[1]{\@@startlink{#1}\@@href}%
\providecommand \@@href[1]{\endgroup#1\@@endlink}%
\providecommand \@sanitize@url [0]{\catcode `\\12\catcode `\$12\catcode
  `\&12\catcode `\#12\catcode `\^12\catcode `\_12\catcode `\%12\relax}%
\providecommand \@@startlink[1]{}%
\providecommand \@@endlink[0]{}%
\providecommand \url  [0]{\begingroup\@sanitize@url \@url }%
\providecommand \@url [1]{\endgroup\@href {#1}{\urlprefix }}%
\providecommand \urlprefix  [0]{URL }%
\providecommand \Eprint [0]{\href }%
\providecommand \doibase [0]{http://dx.doi.org/}%
\providecommand \selectlanguage [0]{\@gobble}%
\providecommand \bibinfo  [0]{\@secondoftwo}%
\providecommand \bibfield  [0]{\@secondoftwo}%
\providecommand \translation [1]{[#1]}%
\providecommand \BibitemOpen [0]{}%
\providecommand \bibitemStop [0]{}%
\providecommand \bibitemNoStop [0]{.\EOS\space}%
\providecommand \EOS [0]{\spacefactor3000\relax}%
\providecommand \BibitemShut  [1]{\csname bibitem#1\endcsname}%
\let\auto@bib@innerbib\@empty
%</preamble>
\bibitem [{\citenamefont {Ahlswede}\ and\ \citenamefont
  {Csisz{\'a}r}(1993)}]{ahlswede1993common}%
  \BibitemOpen
  \bibfield  {author} {\bibinfo {author} {\bibfnamefont {R.}~\bibnamefont
  {Ahlswede}}\ and\ \bibinfo {author} {\bibfnamefont {I.}~\bibnamefont
  {Csisz{\'a}r}},\ }\href@noop {} {\bibfield  {journal} {\bibinfo  {journal}
  {IEEE Transactions on Information Theory}\ }\textbf {\bibinfo {volume} {39}}
  (\bibinfo {year} {1993})}\BibitemShut {NoStop}%
\bibitem [{\citenamefont {Schneier}(1996)}]{schneier1996applied}%
  \BibitemOpen
  \bibfield  {author} {\bibinfo {author} {\bibfnamefont {B.}~\bibnamefont
  {Schneier}},\ }\href@noop {} {\bibfield  {journal} {\bibinfo  {journal} {New
  York.}\ } (\bibinfo {year} {1996})}\BibitemShut {NoStop}%
\bibitem [{\citenamefont {Hillery}\ \emph {et~al.}(1999)\citenamefont
  {Hillery}, \citenamefont {Bu{\v{z}}ek},\ and\ \citenamefont
  {Berthiaume}}]{hillery1999quantum}%
  \BibitemOpen
  \bibfield  {author} {\bibinfo {author} {\bibfnamefont {M.}~\bibnamefont
  {Hillery}}, \bibinfo {author} {\bibfnamefont {V.}~\bibnamefont
  {Bu{\v{z}}ek}}, \ and\ \bibinfo {author} {\bibfnamefont {A.}~\bibnamefont
  {Berthiaume}},\ }\href@noop {} {\bibfield  {journal} {\bibinfo  {journal}
  {Physical Review A}\ }\textbf {\bibinfo {volume} {59}},\ \bibinfo {pages}
  {1829} (\bibinfo {year} {1999})}\BibitemShut {NoStop}%
\bibitem [{\citenamefont {Sen}\ \emph {et~al.}(2003)\citenamefont {Sen},
  \citenamefont {Sen}, \citenamefont {{\.Z}ukowski} \emph
  {et~al.}}]{sen2003unified}%
  \BibitemOpen
  \bibfield  {author} {\bibinfo {author} {\bibfnamefont {A.}~\bibnamefont
  {Sen}}, \bibinfo {author} {\bibfnamefont {U.}~\bibnamefont {Sen}}, \bibinfo
  {author} {\bibfnamefont {M.}~\bibnamefont {{\.Z}ukowski}},  \emph {et~al.},\
  }\href@noop {} {\bibfield  {journal} {\bibinfo  {journal} {Physical Review
  A}\ }\textbf {\bibinfo {volume} {68}},\ \bibinfo {pages} {032309} (\bibinfo
  {year} {2003})}\BibitemShut {NoStop}%
\bibitem [{\citenamefont {Yu}\ \emph {et~al.}(2008)\citenamefont {Yu},
  \citenamefont {Lin},\ and\ \citenamefont {Huang}}]{yu2008quantum}%
  \BibitemOpen
  \bibfield  {author} {\bibinfo {author} {\bibfnamefont {I.-C.}\ \bibnamefont
  {Yu}}, \bibinfo {author} {\bibfnamefont {F.-L.}\ \bibnamefont {Lin}}, \ and\
  \bibinfo {author} {\bibfnamefont {C.-Y.}\ \bibnamefont {Huang}},\ }\href@noop
  {} {\bibfield  {journal} {\bibinfo  {journal} {Physical Review A}\ }\textbf
  {\bibinfo {volume} {78}},\ \bibinfo {pages} {012344} (\bibinfo {year}
  {2008})}\BibitemShut {NoStop}%
\bibitem [{\citenamefont
  {Bandyopadhyay}(2000)}]{bandyopadhyay2000teleportation}%
  \BibitemOpen
  \bibfield  {author} {\bibinfo {author} {\bibfnamefont {S.}~\bibnamefont
  {Bandyopadhyay}},\ }\href@noop {} {\bibfield  {journal} {\bibinfo  {journal}
  {Physical Review A}\ }\textbf {\bibinfo {volume} {62}},\ \bibinfo {pages}
  {012308} (\bibinfo {year} {2000})}\BibitemShut {NoStop}%
\bibitem [{\citenamefont {Nascimento}\ \emph {et~al.}(2001)\citenamefont
  {Nascimento}, \citenamefont {Mueller-Quade},\ and\ \citenamefont
  {Imai}}]{nascimento2001improving}%
  \BibitemOpen
  \bibfield  {author} {\bibinfo {author} {\bibfnamefont {A.~C.}\ \bibnamefont
  {Nascimento}}, \bibinfo {author} {\bibfnamefont {J.}~\bibnamefont
  {Mueller-Quade}}, \ and\ \bibinfo {author} {\bibfnamefont {H.}~\bibnamefont
  {Imai}},\ }\href@noop {} {\bibfield  {journal} {\bibinfo  {journal} {Physical
  Review A}\ }\textbf {\bibinfo {volume} {64}},\ \bibinfo {pages} {042311}
  (\bibinfo {year} {2001})}\BibitemShut {NoStop}%
\bibitem [{\citenamefont {Tyc}\ and\ \citenamefont
  {Sanders}(2002)}]{tyc2002share}%
  \BibitemOpen
  \bibfield  {author} {\bibinfo {author} {\bibfnamefont {T.}~\bibnamefont
  {Tyc}}\ and\ \bibinfo {author} {\bibfnamefont {B.~C.}\ \bibnamefont
  {Sanders}},\ }\href@noop {} {\bibfield  {journal} {\bibinfo  {journal}
  {Physical Review A}\ }\textbf {\bibinfo {volume} {65}},\ \bibinfo {pages}
  {042310} (\bibinfo {year} {2002})}\BibitemShut {NoStop}%
\bibitem [{\citenamefont {Karimipour}\ \emph {et~al.}(2002)\citenamefont
  {Karimipour}, \citenamefont {Bahraminasab},\ and\ \citenamefont
  {Bagherinezhad}}]{karimipour2002entanglement}%
  \BibitemOpen
  \bibfield  {author} {\bibinfo {author} {\bibfnamefont {V.}~\bibnamefont
  {Karimipour}}, \bibinfo {author} {\bibfnamefont {A.}~\bibnamefont
  {Bahraminasab}}, \ and\ \bibinfo {author} {\bibfnamefont {S.}~\bibnamefont
  {Bagherinezhad}},\ }\href@noop {} {\bibfield  {journal} {\bibinfo  {journal}
  {Physical Review A}\ }\textbf {\bibinfo {volume} {65}},\ \bibinfo {pages}
  {042320} (\bibinfo {year} {2002})}\BibitemShut {NoStop}%
\bibitem [{\citenamefont {Bagherinezhad}\ and\ \citenamefont
  {Karimipour}(2003)}]{bagherinezhad2003quantum}%
  \BibitemOpen
  \bibfield  {author} {\bibinfo {author} {\bibfnamefont {S.}~\bibnamefont
  {Bagherinezhad}}\ and\ \bibinfo {author} {\bibfnamefont {V.}~\bibnamefont
  {Karimipour}},\ }\href@noop {} {\bibfield  {journal} {\bibinfo  {journal}
  {Physical Review A}\ }\textbf {\bibinfo {volume} {67}},\ \bibinfo {pages}
  {044302} (\bibinfo {year} {2003})}\BibitemShut {NoStop}%
\bibitem [{\citenamefont {Xiao}\ \emph {et~al.}(2004)\citenamefont {Xiao},
  \citenamefont {Long}, \citenamefont {Deng},\ and\ \citenamefont
  {Pan}}]{xiao2004efficient}%
  \BibitemOpen
  \bibfield  {author} {\bibinfo {author} {\bibfnamefont {L.}~\bibnamefont
  {Xiao}}, \bibinfo {author} {\bibfnamefont {G.~L.}\ \bibnamefont {Long}},
  \bibinfo {author} {\bibfnamefont {F.-G.}\ \bibnamefont {Deng}}, \ and\
  \bibinfo {author} {\bibfnamefont {J.-W.}\ \bibnamefont {Pan}},\ }\href@noop
  {} {\bibfield  {journal} {\bibinfo  {journal} {Physical Review A}\ }\textbf
  {\bibinfo {volume} {69}},\ \bibinfo {pages} {052307} (\bibinfo {year}
  {2004})}\BibitemShut {NoStop}%
\bibitem [{\citenamefont {Fu-Guo}\ \emph {et~al.}(2004)\citenamefont {Fu-Guo},
  \citenamefont {Gui-Lu}, \citenamefont {Yan},\ and\ \citenamefont
  {Li}}]{fu2004increasing}%
  \BibitemOpen
  \bibfield  {author} {\bibinfo {author} {\bibfnamefont {D.}~\bibnamefont
  {Fu-Guo}}, \bibinfo {author} {\bibfnamefont {L.}~\bibnamefont {Gui-Lu}},
  \bibinfo {author} {\bibfnamefont {W.}~\bibnamefont {Yan}}, \ and\ \bibinfo
  {author} {\bibfnamefont {X.}~\bibnamefont {Li}},\ }\href@noop {} {\bibfield
  {journal} {\bibinfo  {journal} {Chinese Physics Letters}\ }\textbf {\bibinfo
  {volume} {21}},\ \bibinfo {pages} {2097} (\bibinfo {year}
  {2004})}\BibitemShut {NoStop}%
\bibitem [{\citenamefont {Li}\ \emph {et~al.}(2004)\citenamefont {Li},
  \citenamefont {Zhang},\ and\ \citenamefont {Peng}}]{li2004multiparty}%
  \BibitemOpen
  \bibfield  {author} {\bibinfo {author} {\bibfnamefont {Y.}~\bibnamefont
  {Li}}, \bibinfo {author} {\bibfnamefont {K.}~\bibnamefont {Zhang}}, \ and\
  \bibinfo {author} {\bibfnamefont {K.}~\bibnamefont {Peng}},\ }\href@noop {}
  {\bibfield  {journal} {\bibinfo  {journal} {Physics Letters A}\ }\textbf
  {\bibinfo {volume} {324}},\ \bibinfo {pages} {420} (\bibinfo {year}
  {2004})}\BibitemShut {NoStop}%
\bibitem [{\citenamefont {Han}\ \emph {et~al.}(2008)\citenamefont {Han},
  \citenamefont {Liu}, \citenamefont {Liu},\ and\ \citenamefont
  {Zhang}}]{han2008multiparty}%
  \BibitemOpen
  \bibfield  {author} {\bibinfo {author} {\bibfnamefont {L.-F.}\ \bibnamefont
  {Han}}, \bibinfo {author} {\bibfnamefont {Y.-M.}\ \bibnamefont {Liu}},
  \bibinfo {author} {\bibfnamefont {J.}~\bibnamefont {Liu}}, \ and\ \bibinfo
  {author} {\bibfnamefont {Z.-J.}\ \bibnamefont {Zhang}},\ }\href@noop {}
  {\bibfield  {journal} {\bibinfo  {journal} {Optics Communications}\ }\textbf
  {\bibinfo {volume} {281}},\ \bibinfo {pages} {2690} (\bibinfo {year}
  {2008})}\BibitemShut {NoStop}%
\bibitem [{\citenamefont {Tittel}\ \emph {et~al.}(2001)\citenamefont {Tittel},
  \citenamefont {Zbinden},\ and\ \citenamefont
  {Gisin}}]{tittel2001experimental}%
  \BibitemOpen
  \bibfield  {author} {\bibinfo {author} {\bibfnamefont {W.}~\bibnamefont
  {Tittel}}, \bibinfo {author} {\bibfnamefont {H.}~\bibnamefont {Zbinden}}, \
  and\ \bibinfo {author} {\bibfnamefont {N.}~\bibnamefont {Gisin}},\
  }\href@noop {} {\bibfield  {journal} {\bibinfo  {journal} {Physical Review
  A}\ }\textbf {\bibinfo {volume} {63}},\ \bibinfo {pages} {042301} (\bibinfo
  {year} {2001})}\BibitemShut {NoStop}%
\bibitem [{\citenamefont {Chen}\ \emph {et~al.}(2005)\citenamefont {Chen},
  \citenamefont {Zhang}, \citenamefont {Zhao}, \citenamefont {Zhou},
  \citenamefont {Lu}, \citenamefont {Peng}, \citenamefont {Yang},\ and\
  \citenamefont {Pan}}]{chen2005experimental}%
  \BibitemOpen
  \bibfield  {author} {\bibinfo {author} {\bibfnamefont {Y.-A.}\ \bibnamefont
  {Chen}}, \bibinfo {author} {\bibfnamefont {A.-N.}\ \bibnamefont {Zhang}},
  \bibinfo {author} {\bibfnamefont {Z.}~\bibnamefont {Zhao}}, \bibinfo {author}
  {\bibfnamefont {X.-Q.}\ \bibnamefont {Zhou}}, \bibinfo {author}
  {\bibfnamefont {C.-Y.}\ \bibnamefont {Lu}}, \bibinfo {author} {\bibfnamefont
  {C.-Z.}\ \bibnamefont {Peng}}, \bibinfo {author} {\bibfnamefont
  {T.}~\bibnamefont {Yang}}, \ and\ \bibinfo {author} {\bibfnamefont {J.-W.}\
  \bibnamefont {Pan}},\ }\href@noop {} {\bibfield  {journal} {\bibinfo
  {journal} {Physical review letters}\ }\textbf {\bibinfo {volume} {95}},\
  \bibinfo {pages} {200502} (\bibinfo {year} {2005})}\BibitemShut {NoStop}%
\bibitem [{\citenamefont {Gaertner}\ \emph {et~al.}(2007)\citenamefont
  {Gaertner}, \citenamefont {Kurtsiefer}, \citenamefont {Bourennane},\ and\
  \citenamefont {Weinfurter}}]{gaertner2007experimental}%
  \BibitemOpen
  \bibfield  {author} {\bibinfo {author} {\bibfnamefont {S.}~\bibnamefont
  {Gaertner}}, \bibinfo {author} {\bibfnamefont {C.}~\bibnamefont
  {Kurtsiefer}}, \bibinfo {author} {\bibfnamefont {M.}~\bibnamefont
  {Bourennane}}, \ and\ \bibinfo {author} {\bibfnamefont {H.}~\bibnamefont
  {Weinfurter}},\ }\href@noop {} {\bibfield  {journal} {\bibinfo  {journal}
  {Physical Review Letters}\ }\textbf {\bibinfo {volume} {98}},\ \bibinfo
  {pages} {020503} (\bibinfo {year} {2007})}\BibitemShut {NoStop}%
\bibitem [{\citenamefont {Guo}\ and\ \citenamefont
  {Guo}(2003)}]{guo2003quantum}%
  \BibitemOpen
  \bibfield  {author} {\bibinfo {author} {\bibfnamefont {G.-P.}\ \bibnamefont
  {Guo}}\ and\ \bibinfo {author} {\bibfnamefont {G.-C.}\ \bibnamefont {Guo}},\
  }\href@noop {} {\bibfield  {journal} {\bibinfo  {journal} {Physics Letters
  A}\ }\textbf {\bibinfo {volume} {310}},\ \bibinfo {pages} {247} (\bibinfo
  {year} {2003})}\BibitemShut {NoStop}%
\bibitem [{\citenamefont {Schmid}\ \emph {et~al.}(2005)\citenamefont {Schmid},
  \citenamefont {Trojek}, \citenamefont {Bourennane}, \citenamefont
  {Kurtsiefer}, \citenamefont {{\.Z}ukowski},\ and\ \citenamefont
  {Weinfurter}}]{schmid2005experimental}%
  \BibitemOpen
  \bibfield  {author} {\bibinfo {author} {\bibfnamefont {C.}~\bibnamefont
  {Schmid}}, \bibinfo {author} {\bibfnamefont {P.}~\bibnamefont {Trojek}},
  \bibinfo {author} {\bibfnamefont {M.}~\bibnamefont {Bourennane}}, \bibinfo
  {author} {\bibfnamefont {C.}~\bibnamefont {Kurtsiefer}}, \bibinfo {author}
  {\bibfnamefont {M.}~\bibnamefont {{\.Z}ukowski}}, \ and\ \bibinfo {author}
  {\bibfnamefont {H.}~\bibnamefont {Weinfurter}},\ }\href@noop {} {\bibfield
  {journal} {\bibinfo  {journal} {Physical review letters}\ }\textbf {\bibinfo
  {volume} {95}},\ \bibinfo {pages} {230505} (\bibinfo {year}
  {2005})}\BibitemShut {NoStop}%
\bibitem [{\citenamefont {He}(2007)}]{he2007comment}%
  \BibitemOpen
  \bibfield  {author} {\bibinfo {author} {\bibfnamefont {G.~P.}\ \bibnamefont
  {He}},\ }\href@noop {} {\bibfield  {journal} {\bibinfo  {journal} {Physical
  review letters}\ }\textbf {\bibinfo {volume} {98}},\ \bibinfo {pages}
  {028901} (\bibinfo {year} {2007})}\BibitemShut {NoStop}%
\bibitem [{\citenamefont {Qin}\ \emph {et~al.}(2008)\citenamefont {Qin},
  \citenamefont {Gao}, \citenamefont {Wen},\ and\ \citenamefont
  {Zhu}}]{qin2008special}%
  \BibitemOpen
  \bibfield  {author} {\bibinfo {author} {\bibfnamefont {S.-J.}\ \bibnamefont
  {Qin}}, \bibinfo {author} {\bibfnamefont {F.}~\bibnamefont {Gao}}, \bibinfo
  {author} {\bibfnamefont {Q.-Y.}\ \bibnamefont {Wen}}, \ and\ \bibinfo
  {author} {\bibfnamefont {F.-C.}\ \bibnamefont {Zhu}},\ }\href@noop {}
  {\bibfield  {journal} {\bibinfo  {journal} {Optics Communications}\ }\textbf
  {\bibinfo {volume} {281}},\ \bibinfo {pages} {5472} (\bibinfo {year}
  {2008})}\BibitemShut {NoStop}%
\bibitem [{\citenamefont {Tavakoli}\ \emph {et~al.}(2015)\citenamefont
  {Tavakoli}, \citenamefont {Herbauts}, \citenamefont {{\.Z}ukowski},\ and\
  \citenamefont {Bourennane}}]{tavakoli2015secret}%
  \BibitemOpen
  \bibfield  {author} {\bibinfo {author} {\bibfnamefont {A.}~\bibnamefont
  {Tavakoli}}, \bibinfo {author} {\bibfnamefont {I.}~\bibnamefont {Herbauts}},
  \bibinfo {author} {\bibfnamefont {M.}~\bibnamefont {{\.Z}ukowski}}, \ and\
  \bibinfo {author} {\bibfnamefont {M.}~\bibnamefont {Bourennane}},\
  }\href@noop {} {\bibfield  {journal} {\bibinfo  {journal} {Physical Review
  A}\ }\textbf {\bibinfo {volume} {92}},\ \bibinfo {pages} {030302} (\bibinfo
  {year} {2015})}\BibitemShut {NoStop}%
\bibitem [{\citenamefont {Karimipour}\ and\ \citenamefont
  {Asoudeh}(2015)}]{karimipour2015quantum}%
  \BibitemOpen
  \bibfield  {author} {\bibinfo {author} {\bibfnamefont {V.}~\bibnamefont
  {Karimipour}}\ and\ \bibinfo {author} {\bibfnamefont {M.}~\bibnamefont
  {Asoudeh}},\ }\href@noop {} {\bibfield  {journal} {\bibinfo  {journal}
  {Physical Review A}\ }\textbf {\bibinfo {volume} {92}},\ \bibinfo {pages}
  {030301} (\bibinfo {year} {2015})}\BibitemShut {NoStop}%
\bibitem [{\citenamefont {Lin}\ \emph {et~al.}(2016)\citenamefont {Lin},
  \citenamefont {Guo}, \citenamefont {Xu}, \citenamefont {Sun},\ and\
  \citenamefont {Liu}}]{lin2016cryptanalysis}%
  \BibitemOpen
  \bibfield  {author} {\bibinfo {author} {\bibfnamefont {S.}~\bibnamefont
  {Lin}}, \bibinfo {author} {\bibfnamefont {G.-D.}\ \bibnamefont {Guo}},
  \bibinfo {author} {\bibfnamefont {Y.-Z.}\ \bibnamefont {Xu}}, \bibinfo
  {author} {\bibfnamefont {Y.}~\bibnamefont {Sun}}, \ and\ \bibinfo {author}
  {\bibfnamefont {X.-F.}\ \bibnamefont {Liu}},\ }\href@noop {} {\bibfield
  {journal} {\bibinfo  {journal} {Physical Review A}\ }\textbf {\bibinfo
  {volume} {93}},\ \bibinfo {pages} {062343} (\bibinfo {year}
  {2016})}\BibitemShut {NoStop}%
\bibitem [{\citenamefont {Chen}\ \emph {et~al.}(2018)\citenamefont {Chen},
  \citenamefont {Tang}, \citenamefont {Xu}, \citenamefont {Dou}, \citenamefont
  {Chen},\ and\ \citenamefont {Yang}}]{chen2018cryptanalysis}%
  \BibitemOpen
  \bibfield  {author} {\bibinfo {author} {\bibfnamefont {X.-B.}\ \bibnamefont
  {Chen}}, \bibinfo {author} {\bibfnamefont {X.}~\bibnamefont {Tang}}, \bibinfo
  {author} {\bibfnamefont {G.}~\bibnamefont {Xu}}, \bibinfo {author}
  {\bibfnamefont {Z.}~\bibnamefont {Dou}}, \bibinfo {author} {\bibfnamefont
  {Y.-L.}\ \bibnamefont {Chen}}, \ and\ \bibinfo {author} {\bibfnamefont
  {Y.-X.}\ \bibnamefont {Yang}},\ }\href@noop {} {\bibfield  {journal}
  {\bibinfo  {journal} {Quantum Information Processing}\ }\textbf {\bibinfo
  {volume} {17}},\ \bibinfo {pages} {225} (\bibinfo {year} {2018})}\BibitemShut
  {NoStop}%
\bibitem [{\citenamefont {Qin}\ and\ \citenamefont
  {Tso}(2019)}]{qin2019efficient}%
  \BibitemOpen
  \bibfield  {author} {\bibinfo {author} {\bibfnamefont {H.}~\bibnamefont
  {Qin}}\ and\ \bibinfo {author} {\bibfnamefont {R.}~\bibnamefont {Tso}},\
  }\href@noop {} {\bibfield  {journal} {\bibinfo  {journal} {Journal of the
  Chinese Institute of Engineers}\ }\textbf {\bibinfo {volume} {42}},\ \bibinfo
  {pages} {143} (\bibinfo {year} {2019})}\BibitemShut {NoStop}%
\bibitem [{\citenamefont {Zhou}\ \emph {et~al.}(2014)\citenamefont {Zhou},
  \citenamefont {Wang}, \citenamefont {Wang},\ and\ \citenamefont
  {Wang}}]{zhou2014implementation}%
  \BibitemOpen
  \bibfield  {author} {\bibinfo {author} {\bibfnamefont {K.-h.}\ \bibnamefont
  {Zhou}}, \bibinfo {author} {\bibfnamefont {Y.}~\bibnamefont {Wang}}, \bibinfo
  {author} {\bibfnamefont {T.-j.}\ \bibnamefont {Wang}}, \ and\ \bibinfo
  {author} {\bibfnamefont {C.}~\bibnamefont {Wang}},\ }\href@noop {} {\bibfield
   {journal} {\bibinfo  {journal} {International Journal of Theoretical
  Physics}\ }\textbf {\bibinfo {volume} {53}},\ \bibinfo {pages} {3927}
  (\bibinfo {year} {2014})}\BibitemShut {NoStop}%
\bibitem [{\citenamefont {Smania}\ \emph {et~al.}(2016)\citenamefont {Smania},
  \citenamefont {Elhassan}, \citenamefont {Tavakoli},\ and\ \citenamefont
  {Bourennane}}]{smania2016experimental}%
  \BibitemOpen
  \bibfield  {author} {\bibinfo {author} {\bibfnamefont {M.}~\bibnamefont
  {Smania}}, \bibinfo {author} {\bibfnamefont {A.~M.}\ \bibnamefont
  {Elhassan}}, \bibinfo {author} {\bibfnamefont {A.}~\bibnamefont {Tavakoli}},
  \ and\ \bibinfo {author} {\bibfnamefont {M.}~\bibnamefont {Bourennane}},\
  }\href@noop {} {\bibfield  {journal} {\bibinfo  {journal} {Npj Quantum
  Information}\ }\textbf {\bibinfo {volume} {2}},\ \bibinfo {pages} {16010}
  (\bibinfo {year} {2016})}\BibitemShut {NoStop}%
\bibitem [{\citenamefont {Marrucci}\ \emph {et~al.}(2006)\citenamefont
  {Marrucci}, \citenamefont {Manzo},\ and\ \citenamefont
  {Paparo}}]{marrucci2006optical}%
  \BibitemOpen
  \bibfield  {author} {\bibinfo {author} {\bibfnamefont {L.}~\bibnamefont
  {Marrucci}}, \bibinfo {author} {\bibfnamefont {C.}~\bibnamefont {Manzo}}, \
  and\ \bibinfo {author} {\bibfnamefont {D.}~\bibnamefont {Paparo}},\
  }\href@noop {} {\bibfield  {journal} {\bibinfo  {journal} {Physical review
  letters}\ }\textbf {\bibinfo {volume} {96}},\ \bibinfo {pages} {163905}
  (\bibinfo {year} {2006})}\BibitemShut {NoStop}%
\bibitem [{\citenamefont {Marrucci}\ \emph {et~al.}(2011)\citenamefont
  {Marrucci}, \citenamefont {Karimi}, \citenamefont {Slussarenko},
  \citenamefont {Piccirillo}, \citenamefont {Santamato}, \citenamefont
  {Nagali},\ and\ \citenamefont {Sciarrino}}]{marrucci2011spin}%
  \BibitemOpen
  \bibfield  {author} {\bibinfo {author} {\bibfnamefont {L.}~\bibnamefont
  {Marrucci}}, \bibinfo {author} {\bibfnamefont {E.}~\bibnamefont {Karimi}},
  \bibinfo {author} {\bibfnamefont {S.}~\bibnamefont {Slussarenko}}, \bibinfo
  {author} {\bibfnamefont {B.}~\bibnamefont {Piccirillo}}, \bibinfo {author}
  {\bibfnamefont {E.}~\bibnamefont {Santamato}}, \bibinfo {author}
  {\bibfnamefont {E.}~\bibnamefont {Nagali}}, \ and\ \bibinfo {author}
  {\bibfnamefont {F.}~\bibnamefont {Sciarrino}},\ }\href@noop {} {\bibfield
  {journal} {\bibinfo  {journal} {Journal of Optics}\ }\textbf {\bibinfo
  {volume} {13}},\ \bibinfo {pages} {064001} (\bibinfo {year}
  {2011})}\BibitemShut {NoStop}%
\bibitem [{\citenamefont {Arriz{\'o}n}\ \emph {et~al.}(2007)\citenamefont
  {Arriz{\'o}n}, \citenamefont {Ruiz}, \citenamefont {Carrada},\ and\
  \citenamefont {Gonz{\'a}lez}}]{arrizon2007pixelated}%
  \BibitemOpen
  \bibfield  {author} {\bibinfo {author} {\bibfnamefont {V.}~\bibnamefont
  {Arriz{\'o}n}}, \bibinfo {author} {\bibfnamefont {U.}~\bibnamefont {Ruiz}},
  \bibinfo {author} {\bibfnamefont {R.}~\bibnamefont {Carrada}}, \ and\
  \bibinfo {author} {\bibfnamefont {L.~A.}\ \bibnamefont {Gonz{\'a}lez}},\
  }\href@noop {} {\bibfield  {journal} {\bibinfo  {journal} {JOSA A}\ }\textbf
  {\bibinfo {volume} {24}},\ \bibinfo {pages} {3500} (\bibinfo {year}
  {2007})}\BibitemShut {NoStop}%
\bibitem [{\citenamefont {Forbes}\ \emph {et~al.}(2016)\citenamefont {Forbes},
  \citenamefont {Dudley},\ and\ \citenamefont {McLaren}}]{forbes2016creation}%
  \BibitemOpen
  \bibfield  {author} {\bibinfo {author} {\bibfnamefont {A.}~\bibnamefont
  {Forbes}}, \bibinfo {author} {\bibfnamefont {A.}~\bibnamefont {Dudley}}, \
  and\ \bibinfo {author} {\bibfnamefont {M.}~\bibnamefont {McLaren}},\
  }\href@noop {} {\bibfield  {journal} {\bibinfo  {journal} {Advances in Optics
  and Photonics}\ }\textbf {\bibinfo {volume} {8}},\ \bibinfo {pages} {200}
  (\bibinfo {year} {2016})}\BibitemShut {NoStop}%
\bibitem [{\citenamefont {Jo}\ \emph {et~al.}(2019)\citenamefont {Jo},
  \citenamefont {Park}, \citenamefont {Lee},\ and\ \citenamefont
  {Son}}]{jo2019efficient}%
  \BibitemOpen
  \bibfield  {author} {\bibinfo {author} {\bibfnamefont {Y.}~\bibnamefont
  {Jo}}, \bibinfo {author} {\bibfnamefont {H.~S.}\ \bibnamefont {Park}},
  \bibinfo {author} {\bibfnamefont {S.-W.}\ \bibnamefont {Lee}}, \ and\
  \bibinfo {author} {\bibfnamefont {W.}~\bibnamefont {Son}},\ }\href@noop {}
  {\bibfield  {journal} {\bibinfo  {journal} {Entropy}\ }\textbf {\bibinfo
  {volume} {21}},\ \bibinfo {pages} {80} (\bibinfo {year} {2019})}\BibitemShut
  {NoStop}%
\bibitem [{\citenamefont {{\.Z}ukowski}\ \emph {et~al.}(1997)\citenamefont
  {{\.Z}ukowski}, \citenamefont {Zeilinger},\ and\ \citenamefont
  {Horne}}]{zukowski1997realizable}%
  \BibitemOpen
  \bibfield  {author} {\bibinfo {author} {\bibfnamefont {M.}~\bibnamefont
  {{\.Z}ukowski}}, \bibinfo {author} {\bibfnamefont {A.}~\bibnamefont
  {Zeilinger}}, \ and\ \bibinfo {author} {\bibfnamefont {M.~A.}\ \bibnamefont
  {Horne}},\ }\href@noop {} {\bibfield  {journal} {\bibinfo  {journal}
  {Physical Review A}\ }\textbf {\bibinfo {volume} {55}},\ \bibinfo {pages}
  {2564} (\bibinfo {year} {1997})}\BibitemShut {NoStop}%
\bibitem [{\citenamefont {Berkhout}\ \emph {et~al.}(2010)\citenamefont
  {Berkhout}, \citenamefont {Lavery}, \citenamefont {Courtial}, \citenamefont
  {Beijersbergen},\ and\ \citenamefont {Padgett}}]{berkhout2010efficient}%
  \BibitemOpen
  \bibfield  {author} {\bibinfo {author} {\bibfnamefont {G.~C.}\ \bibnamefont
  {Berkhout}}, \bibinfo {author} {\bibfnamefont {M.~P.}\ \bibnamefont
  {Lavery}}, \bibinfo {author} {\bibfnamefont {J.}~\bibnamefont {Courtial}},
  \bibinfo {author} {\bibfnamefont {M.~W.}\ \bibnamefont {Beijersbergen}}, \
  and\ \bibinfo {author} {\bibfnamefont {M.~J.}\ \bibnamefont {Padgett}},\
  }\href@noop {} {\bibfield  {journal} {\bibinfo  {journal} {Phys Rev Lett}\
  }\textbf {\bibinfo {volume} {105}},\ \bibinfo {pages} {153601} (\bibinfo
  {year} {2010})}\BibitemShut {NoStop}%
\bibitem [{\citenamefont {Mirhosseini}\ \emph {et~al.}(2013)\citenamefont
  {Mirhosseini}, \citenamefont {Malik}, \citenamefont {Shi},\ and\
  \citenamefont {Boyd}}]{mirhosseini2013efficient}%
  \BibitemOpen
  \bibfield  {author} {\bibinfo {author} {\bibfnamefont {M.}~\bibnamefont
  {Mirhosseini}}, \bibinfo {author} {\bibfnamefont {M.}~\bibnamefont {Malik}},
  \bibinfo {author} {\bibfnamefont {Z.}~\bibnamefont {Shi}}, \ and\ \bibinfo
  {author} {\bibfnamefont {R.~W.}\ \bibnamefont {Boyd}},\ }\href@noop {}
  {\bibfield  {journal} {\bibinfo  {journal} {Nat Commun}\ }\textbf {\bibinfo
  {volume} {4}},\ \bibinfo {pages} {2781} (\bibinfo {year} {2013})}\BibitemShut
  {NoStop}%
\bibitem [{\citenamefont {Ndagano}\ \emph {et~al.}(2017)\citenamefont
  {Ndagano}, \citenamefont {Nape}, \citenamefont {Perez-Garcia}, \citenamefont
  {Scholes}, \citenamefont {Hernandez-Aranda}, \citenamefont {Konrad},
  \citenamefont {Lavery},\ and\ \citenamefont
  {Forbes}}]{ndagano2017deterministic}%
  \BibitemOpen
  \bibfield  {author} {\bibinfo {author} {\bibfnamefont {B.}~\bibnamefont
  {Ndagano}}, \bibinfo {author} {\bibfnamefont {I.}~\bibnamefont {Nape}},
  \bibinfo {author} {\bibfnamefont {B.}~\bibnamefont {Perez-Garcia}}, \bibinfo
  {author} {\bibfnamefont {S.}~\bibnamefont {Scholes}}, \bibinfo {author}
  {\bibfnamefont {R.~I.}\ \bibnamefont {Hernandez-Aranda}}, \bibinfo {author}
  {\bibfnamefont {T.}~\bibnamefont {Konrad}}, \bibinfo {author} {\bibfnamefont
  {M.~P.}\ \bibnamefont {Lavery}}, \ and\ \bibinfo {author} {\bibfnamefont
  {A.}~\bibnamefont {Forbes}},\ }\href@noop {} {\bibfield  {journal} {\bibinfo
  {journal} {Scientific reports}\ }\textbf {\bibinfo {volume} {7}},\ \bibinfo
  {pages} {13882} (\bibinfo {year} {2017})}\BibitemShut {NoStop}%
\bibitem [{\citenamefont {Gisin}\ \emph {et~al.}(2002)\citenamefont {Gisin},
  \citenamefont {Ribordy}, \citenamefont {Tittel},\ and\ \citenamefont
  {Zbinden}}]{gisin2002quantum}%
  \BibitemOpen
  \bibfield  {author} {\bibinfo {author} {\bibfnamefont {N.}~\bibnamefont
  {Gisin}}, \bibinfo {author} {\bibfnamefont {G.}~\bibnamefont {Ribordy}},
  \bibinfo {author} {\bibfnamefont {W.}~\bibnamefont {Tittel}}, \ and\ \bibinfo
  {author} {\bibfnamefont {H.}~\bibnamefont {Zbinden}},\ }\href@noop {}
  {\bibfield  {journal} {\bibinfo  {journal} {Reviews of modern physics}\
  }\textbf {\bibinfo {volume} {74}},\ \bibinfo {pages} {145} (\bibinfo {year}
  {2002})}\BibitemShut {NoStop}%
\bibitem [{\citenamefont {Cerf}\ \emph {et~al.}(2002)\citenamefont {Cerf},
  \citenamefont {Bourennane}, \citenamefont {Karlsson},\ and\ \citenamefont
  {Gisin}}]{cerf2002security}%
  \BibitemOpen
  \bibfield  {author} {\bibinfo {author} {\bibfnamefont {N.~J.}\ \bibnamefont
  {Cerf}}, \bibinfo {author} {\bibfnamefont {M.}~\bibnamefont {Bourennane}},
  \bibinfo {author} {\bibfnamefont {A.}~\bibnamefont {Karlsson}}, \ and\
  \bibinfo {author} {\bibfnamefont {N.}~\bibnamefont {Gisin}},\ }\href@noop {}
  {\bibfield  {journal} {\bibinfo  {journal} {Physical Review Letters}\
  }\textbf {\bibinfo {volume} {88}},\ \bibinfo {pages} {127902} (\bibinfo
  {year} {2002})}\BibitemShut {NoStop}%
\bibitem [{\citenamefont {Nagali}\ \emph {et~al.}(2010)\citenamefont {Nagali},
  \citenamefont {Sansoni}, \citenamefont {Marrucci}, \citenamefont
  {Santamato},\ and\ \citenamefont {Sciarrino}}]{nagali2010experimental}%
  \BibitemOpen
  \bibfield  {author} {\bibinfo {author} {\bibfnamefont {E.}~\bibnamefont
  {Nagali}}, \bibinfo {author} {\bibfnamefont {L.}~\bibnamefont {Sansoni}},
  \bibinfo {author} {\bibfnamefont {L.}~\bibnamefont {Marrucci}}, \bibinfo
  {author} {\bibfnamefont {E.}~\bibnamefont {Santamato}}, \ and\ \bibinfo
  {author} {\bibfnamefont {F.}~\bibnamefont {Sciarrino}},\ }\href@noop {}
  {\bibfield  {journal} {\bibinfo  {journal} {Physical Review A}\ }\textbf
  {\bibinfo {volume} {81}},\ \bibinfo {pages} {052317} (\bibinfo {year}
  {2010})}\BibitemShut {NoStop}%
\bibitem [{\citenamefont {Sheng}\ and\ \citenamefont
  {Deng}(2010)}]{sheng2010deterministic}%
  \BibitemOpen
  \bibfield  {author} {\bibinfo {author} {\bibfnamefont {Y.-B.}\ \bibnamefont
  {Sheng}}\ and\ \bibinfo {author} {\bibfnamefont {F.-G.}\ \bibnamefont
  {Deng}},\ }\href@noop {} {\bibfield  {journal} {\bibinfo  {journal} {Physical
  Review A}\ }\textbf {\bibinfo {volume} {81}},\ \bibinfo {pages} {032307}
  (\bibinfo {year} {2010})}\BibitemShut {NoStop}%
\bibitem [{\citenamefont {Li}(2010)}]{li2010deterministic}%
  \BibitemOpen
  \bibfield  {author} {\bibinfo {author} {\bibfnamefont {X.-H.}\ \bibnamefont
  {Li}},\ }\href@noop {} {\bibfield  {journal} {\bibinfo  {journal} {Physical
  Review A}\ }\textbf {\bibinfo {volume} {82}},\ \bibinfo {pages} {044304}
  (\bibinfo {year} {2010})}\BibitemShut {NoStop}%
\bibitem [{\citenamefont {Barreiro}\ \emph {et~al.}(2008)\citenamefont
  {Barreiro}, \citenamefont {Wei},\ and\ \citenamefont
  {Kwiat}}]{barreiro2008beating}%
  \BibitemOpen
  \bibfield  {author} {\bibinfo {author} {\bibfnamefont {J.~T.}\ \bibnamefont
  {Barreiro}}, \bibinfo {author} {\bibfnamefont {T.-C.}\ \bibnamefont {Wei}}, \
  and\ \bibinfo {author} {\bibfnamefont {P.~G.}\ \bibnamefont {Kwiat}},\
  }\href@noop {} {\bibfield  {journal} {\bibinfo  {journal} {Nature physics}\
  }\textbf {\bibinfo {volume} {4}},\ \bibinfo {pages} {282} (\bibinfo {year}
  {2008})}\BibitemShut {NoStop}%
\bibitem [{\citenamefont {Wang}\ \emph {et~al.}(2009)\citenamefont {Wang},
  \citenamefont {Wang},\ and\ \citenamefont {Long}}]{wang2009doubling}%
  \BibitemOpen
  \bibfield  {author} {\bibinfo {author} {\bibfnamefont {W.-Y.}\ \bibnamefont
  {Wang}}, \bibinfo {author} {\bibfnamefont {C.}~\bibnamefont {Wang}}, \ and\
  \bibinfo {author} {\bibfnamefont {G.-L.}\ \bibnamefont {Long}},\ }\href@noop
  {} {\bibfield  {journal} {\bibinfo  {journal} {International Journal of
  Quantum Information}\ }\textbf {\bibinfo {volume} {7}},\ \bibinfo {pages}
  {529} (\bibinfo {year} {2009})}\BibitemShut {NoStop}%
\bibitem [{\citenamefont {Toninelli}\ \emph {et~al.}(2019)\citenamefont
  {Toninelli}, \citenamefont {Ndagano}, \citenamefont {Vall{\'e}s},
  \citenamefont {Sephton}, \citenamefont {Nape}, \citenamefont {Ambrosio},
  \citenamefont {Capasso}, \citenamefont {Padgett},\ and\ \citenamefont
  {Forbes}}]{toninelli2019concepts}%
  \BibitemOpen
  \bibfield  {author} {\bibinfo {author} {\bibfnamefont {E.}~\bibnamefont
  {Toninelli}}, \bibinfo {author} {\bibfnamefont {B.}~\bibnamefont {Ndagano}},
  \bibinfo {author} {\bibfnamefont {A.}~\bibnamefont {Vall{\'e}s}}, \bibinfo
  {author} {\bibfnamefont {B.}~\bibnamefont {Sephton}}, \bibinfo {author}
  {\bibfnamefont {I.}~\bibnamefont {Nape}}, \bibinfo {author} {\bibfnamefont
  {A.}~\bibnamefont {Ambrosio}}, \bibinfo {author} {\bibfnamefont
  {F.}~\bibnamefont {Capasso}}, \bibinfo {author} {\bibfnamefont {M.~J.}\
  \bibnamefont {Padgett}}, \ and\ \bibinfo {author} {\bibfnamefont
  {A.}~\bibnamefont {Forbes}},\ }\href@noop {} {\bibfield  {journal} {\bibinfo
  {journal} {Advances in Optics and Photonics}\ }\textbf {\bibinfo {volume}
  {11}},\ \bibinfo {pages} {67} (\bibinfo {year} {2019})}\BibitemShut {NoStop}%
\bibitem [{\citenamefont {Cui}\ \emph {et~al.}(2017)\citenamefont {Cui},
  \citenamefont {Su}, \citenamefont {Li},\ and\ \citenamefont
  {Ou}}]{cui2017distribution}%
  \BibitemOpen
  \bibfield  {author} {\bibinfo {author} {\bibfnamefont {L.}~\bibnamefont
  {Cui}}, \bibinfo {author} {\bibfnamefont {J.}~\bibnamefont {Su}}, \bibinfo
  {author} {\bibfnamefont {X.}~\bibnamefont {Li}}, \ and\ \bibinfo {author}
  {\bibfnamefont {Z.}~\bibnamefont {Ou}},\ }\href@noop {} {\bibfield  {journal}
  {\bibinfo  {journal} {Scientific reports}\ }\textbf {\bibinfo {volume} {7}},\
  \bibinfo {pages} {14954} (\bibinfo {year} {2017})}\BibitemShut {NoStop}%
\bibitem [{\citenamefont {Roux}\ and\ \citenamefont
  {Zhang}(2014)}]{roux2014projective}%
  \BibitemOpen
  \bibfield  {author} {\bibinfo {author} {\bibfnamefont {F.~S.}\ \bibnamefont
  {Roux}}\ and\ \bibinfo {author} {\bibfnamefont {Y.}~\bibnamefont {Zhang}},\
  }\href@noop {} {\bibfield  {journal} {\bibinfo  {journal} {Physical Review
  A}\ }\textbf {\bibinfo {volume} {90}},\ \bibinfo {pages} {033835} (\bibinfo
  {year} {2014})}\BibitemShut {NoStop}%
\bibitem [{\citenamefont {Schulze}\ \emph {et~al.}(2013)\citenamefont
  {Schulze}, \citenamefont {Dudley}, \citenamefont {Flamm}, \citenamefont
  {Duparre},\ and\ \citenamefont {Forbes}}]{schulze2013measurement}%
  \BibitemOpen
  \bibfield  {author} {\bibinfo {author} {\bibfnamefont {C.}~\bibnamefont
  {Schulze}}, \bibinfo {author} {\bibfnamefont {A.}~\bibnamefont {Dudley}},
  \bibinfo {author} {\bibfnamefont {D.}~\bibnamefont {Flamm}}, \bibinfo
  {author} {\bibfnamefont {M.}~\bibnamefont {Duparre}}, \ and\ \bibinfo
  {author} {\bibfnamefont {A.}~\bibnamefont {Forbes}},\ }\href@noop {}
  {\bibfield  {journal} {\bibinfo  {journal} {New Journal of Physics}\ }\textbf
  {\bibinfo {volume} {15}},\ \bibinfo {pages} {073025} (\bibinfo {year}
  {2013})}\BibitemShut {NoStop}%
\end{thebibliography}%



\section{Supplementary material}

\subsection*{Modal decomposition}
We employed modal decomposition for performing the inner-product measurements, i.e mode projections. This technique is used for performing optical projective measurements in the quantum and classical regime \cite{roux2014projective}. 

Firstly, to perform the modal overlap between the normalised spatial modes $\psi(\textbf{r})$ and  $\phi(\textbf{r})$, we simply compute the inner-product 
%%%%%%%%%%%%%%%%%%%%%%%%%%%%%%%%%%%%%%%%%%%%%%%%
\begin{align}
c &= \braket{\phi | \psi} \nonumber\\ 
       &= \iint \phi^{*}(\textbf{r}) \psi(\textbf{r}) \ d^{2}r.
\label{eqn:Overlap} 
\end{align}
%%%%%%%%%%%%%%%%%%%%%%%%%%%%%%%%%%%%%%%%%%%%%%%%
Here $r = (x,y)$ while $|c|^2$ is the overlap probability determining the correlation between the two modes. Accordingly, any arbitrary input field, $\psi(\textbf{r})$, can be correlated with a second mode $\phi(\textbf{r})$ where $|c|^2$=1 for a high correlation, meaning the modes are equivalent and $|c|^2$=0 for no correlation meaning that the modes are orthogonal. Optically, $\phi(\textbf{r})$ can be a match filter \cite{schulze2013measurement} in the form of a hologram encoded on an SLM. As such the overlap probability $|c| ^2$ can be obtained by taking the Fourier transform (using a Fourier lens \cite{schulze2013measurement}) of the product $\phi^{*}(\textbf{r}) \psi(\textbf{r})$, which is the output mode after the match filter, hence yielding the state:
%%%%%%%%%%%%%%%%%%%%%%%%%%%%%%%%%%%%%%%%%%%%%%%%
\begin{equation}
A(k_{x},k_{y}) = \iint \phi^*(x,y)\psi(x,y) 
e^{-i(k_{x}x + k_{y}y)} \ dxdy
\label{eqn:FT}
\end{equation}
%%%%%%%%%%%%%%%%%%%%%%%%%%%%%%%%%%%%%%%%%%%%%%%%
\noindent where $k_{x}, k_{y}$ are transverse wave vectors in Cartesian coordinates. Evaluating the on-axis point ($k_{x},k_{y}$) = $(0,0)$, results in  Eq. \ref{eqn:Overlap}. Therefore 
%%%%%%%%%%%%%%%%%%%%%%%%%%%%%%%%%%%%%%%%%%%%%%%%
\begin{equation}
A(0,0) = \iint \phi^{*}(\textbf{r})\psi(\textbf{r}) \ d^{2}r = c
\label{eqn:OPTproduct}
\end{equation}
%%%%%%%%%%%%%%%%%%%%%%%%%%%%%%%%%%%%%%%%%%%%%%%%
results in the intensity at the field center, $I(0,0)=|A(0,0)|^2$, being the modal overlap weighting (equivalently detection probability) $|c|^2$. We used this technique to perform our optical projective measurements.
%%%%%%%%%%%%%%%%%%%%%%%%%%%%%%%%%%%%%%%%%%%%%%%%%%%%%%%%%%%%%%%%%%%%%%%%%%%%%%%%%%%%%%%%%%%%%%%%%%%%%%%%%%%%%%%%%%%%%%%%%%%%%%%%%%%%%%%%%%%%%%%%%%%%%%%%%%%%%%%
\end{document}